\documentclass[a4paper,12pt]{article}

\usepackage{graphicx}
\usepackage{hyperref}
\usepackage{epsfig}
\usepackage{amssymb}
\usepackage{setspace}

\usepackage{amsmath}
\usepackage{amssymb}
\usepackage{graphicx}
\usepackage{geometry}
\usepackage{subfigure}
\usepackage{url}
\usepackage[punctsep]{collref}

\newcommand{\be}{\begin{equation}}
\newcommand{\ee}{\end{equation}}
\newcommand{\br}{\begin{eqnarray}}
\newcommand{\bea}{\begin{eqnarray}}
\newcommand{\eea}{\end{eqnarray}}
\newcommand{\er}{\end{eqnarray}}
\newcommand{\ba}{\begin{array}}
\newcommand{\ea}{\end{array}}
\newcommand{\bi}{\begin{itemize}}
\newcommand{\ei}{\end{itemize}}
\newcommand{\bn}{\begin{enumerate}}
\newcommand{\en}{\end{enumerate}}
\newcommand{\bc}{\begin{center}}
\newcommand{\ec}{\end{center}}

 \def\({\left(}
 \def\){\right)}
 \def\[{\left[}
 \def\]{\right]}

\def\mHu{{m_{h_u}}}
 \def\mHd{{m_{h_d}}}
 \def\mHu2{{m_{h_u}^2}}
 \def\mHd2{{m_{h_d}^2}}

\def\gappeq{\mathrel{\rlap {\raise.5ex\hbox{$>$}}
{\lower.5ex\hbox{$\sim$}}}}

\def\lappeq{\mathrel{\rlap{\raise.5ex\hbox{$<$}}
{\lower.5ex\hbox{$\sim$}}}}

\begin{document}
\pagestyle{empty}
%\begin{flushright}
%CERN-PH-TH-2012-078 \\
%\end{flushright}
%\vspace*{10mm}
\begin{center}
{\LARGE {\bf  Implications of effective axial-vector 
coupling \\
\vspace{0.3cm}
of gluon for  $t \bar t$ spin polarizations at the LHC}} \\
\vspace*{1.5cm}
{\large
 {\bf E. Gabrielli$^{{a,b,}}$\footnote{
On leave of absence from Dipartimento di Fisica  Universit\`a di 
Trieste, Strada Costiera 11, I-34151 Trieste}
}, 
 {\bf A.~Racioppi$^{a}$},
  {\bf M.~Raidal$^{a,c}$}, {\bf H.~Veerm\"ae$^{a,c}$}}
\vspace{0.3cm}

{\it
 (a) NICPB, R\"avala 10, Tallinn 10143, Estonia}  \\[1mm]
{\it
 (b) INFN sezione di Trieste, via Valerio 2, I-34127 Trieste, Italy}  \\[1mm]
{\it  (c) Institute of Physics, 
University of Tartu, Riia 142, 51014 Tartu,Estonia} \\[1mm]

\vspace*{3cm}
{\bf ABSTRACT} \\
\end{center}
\vspace*{5mm}

\noindent
We analyze the impact of effective axial-vector  coupling of the gluon  on spin polarization observables 
in $t\bar{t}$ pair production at the LHC. Working at leading order in QCD, we compute the $t\bar{t}$ spin-correlation and left-right spin 
asymmetry coefficients in the helicity basis in the laboratory frame as  functions of the new physics scale $\Lambda$ 
associated with this coupling. We found that the $t\bar{t}$ invariant mass dependent asymmetries are more sensitive to the scale 
$\Lambda$ than the corresponding inclusive ones, in particular when suitable cuts selecting high $t\bar{t}$ invariant mass regions
are imposed. In the context  of this scenario, we show that the LHC has potential either to confirm or to rule out the Tevatron FB top 
asymmetry anomaly  by analyzing the $t\bar{t}$ spin-correlation and left-right polarization asymmetries.
On the other hand, stringent lower bound on the new physics scale $\Lambda$ can be set in this scenario
if no significant deviations from the SM predictions for those observables will be measured.

\vfill\eject
%\pagestyle{empty}
%\clearpage\mbox{}\clearpage

\setcounter{page}{1}
\pagestyle{plain}

%INSERT YOUR TEXT HERE

\section{Introduction}

Top-quark physics is undoubtedly the best framework where to
study polarized processes at the level of fundamental interactions
\cite{Beneke:2000hk}. Discovered at Tevatron in 1995 \cite{Abachi:1994td},
\cite{Abe:1995hr}  and
copiously produced both at the Tevatron and the Large Hadron Collider (LHC)
\cite{Schilling:2012dx}, 
the top-quark is the 
heaviest elementary fermion with the measured mass of 
$m_t=172.9\pm 0.6 \pm 0.9$ GeV and decay width of
$\Gamma_t=2.0^{+0.7}_{-0.6}$ GeV \cite{Beringer:1900zz}. Since the top
life-time is shorter than the characteristic 
hadronization time-scale 
$\sim 1/\Lambda_{\rm QCD}$,
with $\Lambda_{\rm QCD}$ the
characteristic QCD energy scale, this guarantees that it
will always decay before hadronizing. 
Indeed, the top decay, which is dominated by the weak decay channel
$t\to W b$, is
expected to occur before its spin is flipped by strong interactions. This
ensures that top spin-polarization
at production level will be fully transferred to its decay products. 
Then, the spin of the top-quark can be accessed by measuring the angular 
distributions of the final state decay products. The
QCD corrections to the $t\bar{t}$ pair production at hadron colliders can be 
safely computed at
high orders in perturbation theory \cite{Nason:1987xz,Nason:1989zy,Beenakker:1988bq,Beenakker:1990maa,Mangano:1991jk,Laenen:1993xr,Frixione:1995fj,Kidonakis:1997gm,Bonciani:1998vc,Kidonakis:2001nj,Cacciari:2008zb,Moch:2008qy,Kidonakis:2009mx,Ahrens:2011px,Kidonakis:2011jg}, which allow us to determine the top-quark
polarization with high accuracy.

An interesting observable that can be measured with high precision 
at hadron colliders is the spin-correlation in the 
$t\bar{t}$ pair production. This observable was analyzed in
\cite{
Kuhn:1983ix,Barger:1988jj,Kane:1991bg,Arens:1992fg,Mahlon:1995zn,Stelzer:1995gc,
Brandenburg:1996df,Chang:1995ay,Bernreuther:1995cx,Dharmaratna:1996xd,Mahlon:1997uc,Uwer:2004vp} at the leading order (LO),
and is now known at the next to leading order (NLO) 
\cite{Bernreuther:2001rq,Bernreuther:2004jv} in the strong coupling
in QCD, while more recently the NLO weak corrections 
\cite{Bernreuther:2010ny} have been included. 
The standard model (SM) predicts that spins of the
top- and antitop-quarks are strongly correlated, 
which is just a consequence of the partonic $t\bar{t}$
production mechanisms at hadron colliders.
The tree-level partonic processes, contributing to 
the $t\bar{t}$ production at the LO in 
QCD, are the quark-antiquark and gluon-gluon annihilation processes,
namely $q\bar{q}\to t \bar{t}$ and $g g \to t \bar{t}$ respectively
\cite{Beneke:2000hk}. While at Tevatron the 
first mechanism dominates, the second one is the leading $t\bar{t}$ 
production mechanism at the LHC, 
contributing to almost 90\% of the $pp \to t\bar{t} X$ 
total cross section. Therefore, at low $t\bar{t}$ invariant mass, 
the top- and antitop-quarks are mainly 
produced at the LHC experiments in the left-left and right-right helicity configurations, due
to the spin-1 nature of gluons \cite{Mahlon:2010gw}.
The different production mechanisms and collision energies in the Tevatron and the LHC
 make the measurements of $t\bar{t}$ spin correlations in those experiments complementary to each other.
This observable is 
also a very sensitive probe of  new physics scenarios
that contribute to the partonic $t\bar{t}$ production mechanisms, 
whilst keeping the $t\bar{t}$ production cross section at hadron colliders
within experimental and theoretical bounds \cite{Kane:1991bg,
Frederix:2007gi,Arai:2009cp,Cheung:1996kc,Liu:2010zze,Biswal:2012dr}.

Both CDF \cite{Aaltonen:2010nz}  and D0 \cite{Abazov:2011ka,Abazov:2011qu,Abazov:2011gi} Collaborations at Tevatron 
have performed measurements of the $t\bar{t}$ spin-correlation which, within experimental errors,
are in agreement with the NLO SM predictions. 
 In particular, 
D0 Collaboration  has reported an evidence for the spin-correlation in 
$t\bar{t}$ with a significance of 3.1$\sigma$  \cite{Abazov:2011gi}.
From the LHC, the ATLAS \cite{ATLAS:2012ao} 
and CMS \cite{CMS:2102} 
Collaborations  have 
recently analyzed the $t\bar{t}$ spin correlation 
by analyzing $\sqrt{s}=7$ TeV data corresponding to 
an integrated luminosity of about 2.1 ${\rm fb}^{-1}$ and  5 ${\rm fb}^{-1}$, respectively. ATLAS 
has excluded the hypothesis of zero spin-correlation with a
significance of 5.1 $\sigma$ \cite{ATLAS:2012ao}, while the CMS 
has only reported a 2.9 $\sigma$ evidence \cite{CMS:2102}. 
Within experimental errors both measurements are
consistent with the NLO SM predictions  \cite{Bernreuther:2010ny}.

However, despite the good agreement between the SM and data
in the top-quark sector, the 3$\sigma$ excess  in the $t\bar{t}$ charge or forward-backward (FB) asymmetry with respect to the SM predictions 
\cite{Kuhn:1998jr,Kuhn:1998kw,Bowen:2005ap},
observed at Tevatron by the CDF \cite{Aaltonen:2011kc} and D0 
\cite{Abazov:2011rq} Collaborations, still needs to be clarified.
The intriguing property of this anomalous measurement  is that 
the charge asymmetry increases with 
the  $t\bar{t}$ invariant mass. At the same time the 
measured $t\bar{t}$ production cross section is consistent, within experimental
errors, with the SM prediction \cite{
Nason:1987xz,Nason:1989zy,Beenakker:1988bq,Beenakker:1990maa,Mangano:1991jk,Laenen:1993xr,Frixione:1995fj,Kidonakis:1997gm,Bonciani:1998vc,Kidonakis:2001nj, Kidonakis:2009mx, Kidonakis:2011jg}, both at Tevatron 
\cite{Aaltonen:2010bs,Abazov:2009si} and at the LHC \cite{ATLAS-CONF-2011-140,CMS-PAS-TOP-11-007}.

Numerous new physics models have been proposed to explain this excess of events.
Most of them predict the existence of new particles 
that have parity violating interactions with quarks. 
In particular,  models with flavor dependent axigluons 
\cite{Antunano:2007da,Ferrario:2008wm,Ferrario:2009bz,Frampton:2009rk,Chivukula:2010fk,Bai:2011ed,wang:2011taa,Haisch:2011up,AguilarSaavedra:2011ci,Krnjaic:2011ub,AguilarSaavedra:2011ck}, 
flavor-changing $Z^{\prime}$ interactions 
\cite{Jung:2009jz,Xiao:2010hm,Cao:2011ew,Berger:2011ua,AguilarSaavedra:2011zy,
Jung:2011ue,Duraisamy:2011pt,Ko:2011vd}  or $W^{\prime}$
\cite{Cheung:2009ch,Cheung:2011qa,Bhattacherjee:2011nr,Barger:2011ih,
Craig:2011an,Chen:2011mga,Yan:2011tf,Knapen:2011hu}
 have been suggested. However, in order to reduce the tension with
the SM prediction, these
new particles should be relatively light. In particular, the new particle masses
span from a few hundred GeV in the case of weakly interacting particles
up to 1-2 TeV for the strongly interacting ones, such as the axigluons.
Some of these models are now strongly constrained by negative searches of
new heavy particles, like flavor-changing couplings to top quark
\cite{Chatrchyan:2011dk}, and contact terms interactions 
\cite{Khachatryan:2011as,Aad:2011aj} at the LHC.

In Ref.\cite{Gabrielli:2011jf} it was shown that 
the Tevatron anomaly could be explained by introducing a universal
effective axial-vector coupling of the gluon with quarks.
This effective coupling arises also in the SM, being induced at one-loop by weak radiative 
corrections \cite{Bohm:1986rj}. 
However, it is too small to account for the Tevatron anomaly.
Although such an anomalous coupling could have different new physics (NP) origins, its
main feature is that it naturally predicts 
the correct sign for the asymmetry and does not necessarily 
require new light resonances. As shown in \cite{Gabrielli:2011jf}, 
the characteristic 
new physics scale $\Lambda$ associated to this coupling  should lie in a narrow range $\Lambda \simeq 1-1.3$ TeV. 
 This range has been found to correctly reproduce
the Tevatron anomaly on top-quark charge asymmetry, while
the lower bound on $\Lambda > 1$ TeV 
comes mainly from requiring conservative constraints on the total cross section 
of top-quark pair production at Tevatron.

More recently, in \cite{Gabrielli:2011zw}, the implications of this scenario has been analyzed 
for various top-quark charge asymmetries at the LHC 
\cite{Kuhn:2011ri,Rodrigo:2010gm}. 
In particular,  it was shown that the
LHC with 7-8 TeV center of mass energy 
has the potential either to rule out or 
strongly constrain this scenario \cite{Gabrielli:2011zw}. 
This would require to analyze the cut-dependent charge asymmetries 
at different invariant masses of the $t\bar{t}$ system, 
as a function of $t\bar{t}$ invariant mass $m_{tt}$.  Large deviations 
from the SM prediction are indeed expected to appear in regions
of $m_{tt}$ close to the $\Lambda$ scale. On the other hand, 
when inclusive observables in the kinematic range of $m_{tt}$ are considered, 
the new physics contribution for a scale $\Lambda > 1$ TeV turns out to be 
smaller than the SM contribution. This picture is consistent with
present LHC measurements of top-antitop charge asymmetries 
\cite{CMS-PAS-TOP-11-014,Chatrchyan:2011hk,ATLAS-CONF-2011-106}, which 
are inclusive in $m_{tt}$ and consistent  with the SM prediction \cite{Kuhn:2011ri}.

The aim of the present work is to extend the analysis of
\cite{Gabrielli:2011jf,Gabrielli:2011zw}, by computing 
the effect of this scenario on the $t\bar{t}$
spin observables that can be measured at the LHC.
In particular, we will analyze 
the spin-correlation and the left-right (LR) polarization asymmetry
\cite{Li:1997ae,Li:1997gh,Kao:1994rn,Sullivan:1996ry,Kao:1999kj,Cao:2010nw}
 in 
the laboratory frame, as a function of the new physics scale $\Lambda$.
Will show that these observable, when computed on the $t\bar{t}$
high invariant mass ($m_{tt}$) regions,  are
very sensitive to a scale $\Lambda$ in the TeV range.

Regarding the recent ATLAS \cite{ATLAS:2012ao} and CMS \cite{CMS:2102}
measurements of $t\bar{t}$ spin-correlations,
a direct comparison with these results is not possible in the approach
of \cite{Gabrielli:2011jf,Gabrielli:2011zw},
where a low energy parametrization of the effective gluon 
axial-vector vertex has been adopted.
Indeed, the measurements in \cite{ATLAS:2012ao},
\cite{CMS:2102} are inclusive in the  $m_{tt}$ invariant mass, while the 
low energy approximation, used in \cite{Gabrielli:2011jf,Gabrielli:2011zw} 
to parametrize this effective vertex,
breaks down for values of $m_{tt} > \Lambda$, due to the breaking of
perturbative unitarity. However, this is an artefact of the low energy approximation, since 
the effective gluon axial-vector coupling, being related to 
an operator of dimension 4, has a momentum dependence which is 
valid at any energy scale $m_{tt} > 2 m_t$. 
Indeed, this is the case, for instance, of the SM
where the gluon axial-vector coupling is generated at one-loop 
by the electroweak corrections \cite{Bohm:1986rj}.

In order to circumvent this problem, and extend the predictions to the
kinematic regions
$m_{tt} > \Lambda$, we will assume a particular shape of the form factor 
that would respect unitarity and perturbation theory. 
In particular, we will assume that this effective coupling tends to a 
cut-off in the asymptotic limit $m_{tt} \gg \Lambda$, while it satisfies the 
low energy limit required by QCD Ward identities.
In this way, a direct comparison with the results in 
\cite{ATLAS:2012ao}, \cite{CMS:2102} would be possible, although
at the price of introducing a new free parameter 
and a particular shape of the form factor.
The purpose of this test is to check that 
 the $m_{tt}$ inclusive top spin correlation observables 
are mainly dominated by the
kinematic regions $m_{tt}< \Lambda \sim 1$ TeV, and therefore they are
 not very sensitive to cut-off values of order ${\cal O}(1)$
and to the shape of the form factor. 
In particular, we will show
that in the context of this scenario, values of $\Lambda>1$ TeV are 
still consistent, within two standard deviations, with the recent ATLAS and 
CMS recent measurements.
These results suggest that
a  dedicated experimental analysis at the LHC is needed that studies 
the $t\bar{t}$ spin correlation dependence on the $t \bar t$ invariant mass $m_{tt}$  in order to either confirm or strongly constrain this scenario.

The paper is organized as follows. In Sec. II we
review the theoretical framework  and 
provide the analytical expressions for contribution of  
the effective gluon axial-vector coupling to 
the polarized $q\bar{q}\to t\bar{t}$ and $gg \to t\bar{t}$ 
total cross sections. 
In Sec. III we study the effects of this scenario on the $t\bar{t}$
spin-correlation and left-right top-quark asymmetry at the LHC.
Finally, in Sec. IV we give our conclusions. In Appendix we report the
analytical expressions for the corresponding amplitudes in the helicity
basis, and their square moduli given for all possible final spin 
configurations, for the $q\bar{q}\to t\bar{t}$ and $gg \to t\bar{t}$ processes.

\section{Polarized processes}

\subsection{Theoretical framework}
The most general effective vertex $\Gamma^{a\, \mu}(q^2,M)$ for a quark-gluon
interaction, in momentum space, 
containing the contribution of lowest dimensional operators, and
compatible with gauge-, CP-, and Lorentz-invariance, is
\cite{Gabrielli:2011jf} \bea
\Gamma^{a\, \mu}(q^2,M)
&=& -ig_sT^a\Big\{
\gamma^\mu\left(1+ g_V(q^2,M)
+ \gamma_5 g_A(q^2,M)\right)  
\nonumber \\
&+&
 g_P(q^2,M)q^{\mu}\gamma_5 +
g_M(q^2,M) \sigma^{\mu\nu} q^{\nu} \Big\} ,
\label{vertex}
\eea
where $g_S$ is the strong coupling constant, and $T^a$ are the color matrices. 
In general, the $g_{V,A,P,M}$ form factors depend by a characteristic 
energy scale $M$, typically the largest mass 
scale running in the loops, and by $q^2$ which is the invariant 
momentum-squared carried by the gluon.  
The $g_{V,A,P,M}$ form factors can also depend by the quark flavor.
In the following, we will introduce the dependence on the flavor 
in the form factors when required.

All the effective
couplings appearing in Eq.(\ref{vertex}) arise 
also in the SM at the one-loop level due to the weak corrections \cite{Bohm:1986rj}.
The corresponding scale
$M$ in that case is connected to the electroweak (EW) scale, being 
induced by the exchange of W and Z weak bosons in the loop.

The SM contribution to the parity-violating $g_A,g_P$ couplings, which
is a typical EW correction to the gluon-quark vertex,
 is expected to be small and  cannot explain the Tevatron anomaly \cite{Gabrielli:2011jf}. Recently,  the NLO weak corrections to the
forward-backward  and charge asymmetry at Tevatron and LHC has been
computed \cite{Hollik:2011ps} and their effect account for a few percent. 

Finally, the last term in Eq.(\ref{vertex}) is the contribution of the chromomagnetic dipole operator (with $g_M$ the corresponding form factor), that may
affect the total cross section \cite{Haberl:1995ek,Hioki:2009hm}
but does not significantly contribute to 
the top-quark FB asymmetry \cite{Blum:2011up}.

The QCD gauge invariance requires that
\bea
q_{\mu} \bar{U}_f(p_1)\, \Gamma^{a\, \mu}(q^ 2,M)\, U_f(p_2)&=&0\, ,
\eea
where in the above equation $q=p_1-p_2$ and 
the external bi-spinors $U_f(p_{1,2})$ associated to the quark flavor $f$ 
in momentum space are understood to be on-shell.
Model independently, this condition implies the following Ward identity
\bea 2m_Q
g_A(q^2,M)=q^2 g_P(q^2,M),
\eea 
thus 
\bea 
\lim_{q^2\to 0}
g_{A,V}(q^2,M)=0\, ,
\label{ginv} 
\eea 
since no $1/q^2$
singularities are present in $g_P$. 
Notice that the Ward identity in Eq.(\ref{ginv}) is exact and free from 
any anomaly contribution, since the vector-axial coupling is an effective 
vertex and the fundamental theory (QCD) is anomaly free.
For a more detailed discussion regarding the origin of the form factors
$g_{A,P}(q^2,M)$ associated to the quark of flavor $f$, see Refs.\cite{
Gabrielli:2011jf},\cite{Gabrielli:2011zw}.

 In Ref.\cite{Gabrielli:2011jf} we found that
the magnitude of $g_A$, necessary to explain the Tevatron
$A_{FB}^t$ anomaly, is not compatible with the condition $g_A \sim
g_V$, since $g_V$ is strongly constrained by the measurements on
the $p \bar p \to t \bar{t}$ cross section, which are in good
agreement with the SM prediction.
Then, following the same approach as in \cite{Gabrielli:2011jf}, from
now on, we will neglect the contribution of the vectorial form
factor $g_V(q^2,M)$ in Eq.~(\ref{vertex}), and consider only NP
scenarios that generate $g_A$ with the hierarchy $g_V \ll g_A$. In
the limit of $q^2\ll M^2$, it is useful to parametrize the
axial-vector form factor as 
\bea 
g_A(q^2,M)=\frac{q^2}{\Lambda^2}
F(q^2,\Lambda)\, , 
\label{gA} \eea 
where  we absorb the NP
coupling $\alpha_{NP}$ and loop factor into the NP scale,
$\Lambda^2=M^2/(4\pi \alpha_{NP}).$
Because of the breaking of
conformal invariance, induced by renormalization, we expect
\cite{Raidal:1997hq} $F(q^2,\Lambda)$ to contain also logarithm
terms  $\log(q^2/\Lambda^2).$ This could give a large log
enhancement in the case of $|q^2|\ll \Lambda^2$. In general, the
form factor $F(q^2,\Lambda)$ could also develop an imaginary part
for $q^2>0$. In perturbation theory, this is related to the
absorptive part of the loop diagram generating $g_A$, when $|q^2|$
is above the threshold of some specific particles pair production.

Below, we will analyze the contribution of the axial-vector
$g_A$ anomalous coupling,
as defined in Eq.~(\ref{vertex}), to the polarized partonic cross sections
for $t\bar t$ pair production at the LHC,
related to the processes  $q\bar{q}\to t\bar t$ and $g g\to t\bar t$.
In order to give more general results, we will introduce in the following
the dependence of the quark flavor $f=q,t$ in the effective gluon axial-vector
coupling $g^f_A$, where symbols $q$ and $t$ stand for a generic light quark and 
top-quark respectively.

\subsection{Polarized $q\bar{q}\to t\bar t$ process}
\begin{figure}[t]
\begin{center}
\includegraphics[width=0.60\textwidth]{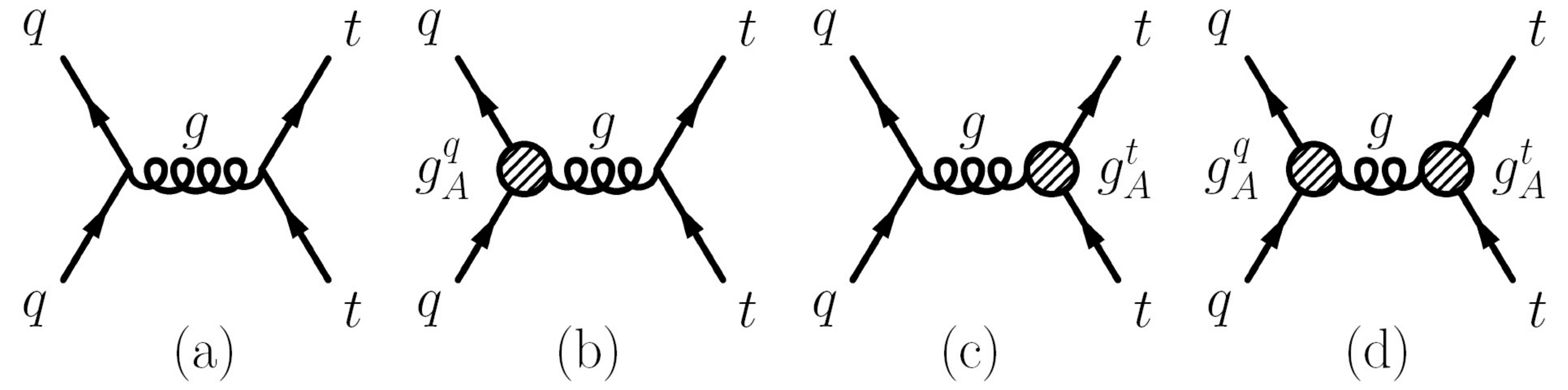}
\vspace{-0.3cm}
\caption{Feynman diagrams (a)-(d)
for the $q\bar{q} \to t \bar t$ process, with
the contribution of the gluon effective axial-vector coupling to light 
quarks ($g^q_A$) and top-quark ($g^t_A$).}
\label{fig:qqtt}
\end{center}
\end{figure}

Let us consider the tree-level scattering
\bea
q(p_1) \bar{q}(p_2)\to t(p_3)\bar t(p_4)\, ,
\label{qqprocess}
\eea
where $p_{1-4}$ are the corresponding particles momenta and $q$ stands for
a light quark.
The Feynman diagrams (a)-(d) relative to
$q\bar q \to t\bar t$, including the axial-vector coupling,
are shown in Fig. \ref{fig:qqtt}. 
According to Eq.~(\ref{vertex}), supplemented
by the Ward identity in Eq.~(\ref{ginv}),
the Feynman rule $\Gamma_A^{a~ \mu}$,
corresponding to the effective axial-vector gluon couplings to quarks $q$ is
\bea
\Gamma_A^{a~ \mu}=
i g^q_A\,
T^a\left(\gamma_{\mu}\gamma_5 -2 q_{\mu}\frac{m_q}{q^2}\gamma_5\right)\, ,
\label{Frule}
\eea
where $q_{\mu}$ is the gluon momentum entering the vertex, $m_q$ is the
quark mass, and $T^a$ the color matrix. From now on, to lighten the notation,
we will omit the $q^2$ and any other mass scale 
dependence in the $g^q_A$ form factors, unless specified.

Below we will give the analytical expressions for the polarized total 
cross sections for the process $q\bar{q}\to t\bar{t}$, in the helicity 
basis and in the $q\bar{q}$ center of mass frame or zero momentum frame (ZMF).
In Appendix we will provide the analytical expressions 
for the corresponding amplitudes in the helicity
basis in the ZMF, and their square moduli given for all possible final spin 
configurations.

The results for the polarized total cross sections are the following
\bea
\sigma^{q\bar{q}}_{LL}(\hat{s})&=&\frac{2 \pi \alpha_S^2}{27 \hat{s}} 
\beta \rho\left(1+|g_A^q|^2\right),
\nonumber\\
\sigma^{q\bar{q}}_{LR}(\hat{s})&=&\frac{4 \pi \alpha_S^2}{27 \hat{s}} \left(1+|g_A^q|^2\right)
\left( 2 {\rm Re}[g_A^t](1-\rho)+\beta\left(1+|g_A^t|^2(1-\rho)\right)
\right),
\nonumber\\
\sigma^{q\bar{q}}_{RR}(\hat{s})&=&\sigma^{q\bar{q}}_{LL}(\hat{s}),
\nonumber\\
\sigma^{q\bar{q}}_{RL}(\hat{s})&=&\sigma^{q\bar{q}}_{LR}(\hat{s}) \Big\{ 
 {\rm Re}[g_A^t] \to - {\rm Re}[g_A^t] \Big\},
\label{crosspolqq}
\eea
where we neglect the mass of the initial light quarks, 
$\beta=\sqrt{1-\rho}$, with $\rho=4m_t^2/\hat{s}$, and
$\hat{s}=(p_1+p_2)^2$. The total sum over polarization is in agreement
with the unpolarized corresponding result in 
\cite{Gabrielli:2011jf,Gabrielli:2011zw}.

As we can see from the results in  Eq.(\ref{crosspolqq}), the
left-right (LR) symmetry, obtained by the simultaneous exchange of 
left-handed with right-handed top-quark polarizations, 
is broken at the tree-level 
by the presence of the axial-vector coupling of the gluon.
On the other hand, in pure QCD the top-quark LR symmetry 
remains exact at any order in perturbation theory due to the parity conservation
of strong interactions, while it is
broken at one-loop by the effect of weak radiative corrections
\cite{Li:1997ae,Li:1997gh,Kao:1994rn,Sullivan:1996ry,Kao:1999kj,Cao:2010nw}.
On the other hand, the vector-axial coupling of gluon can induce the LR 
symmetry-breaking on top-quark polarizations at the tree-level.
This suggests that any observable based on the LR asymmetry of top-quark 
polarizations turns out to be a very sensitive probe of this scenario.

\subsection{Polarized $gg\to t\bar t$ process}
\begin{figure}[t]
\begin{center}
\includegraphics[width=0.60\textwidth]{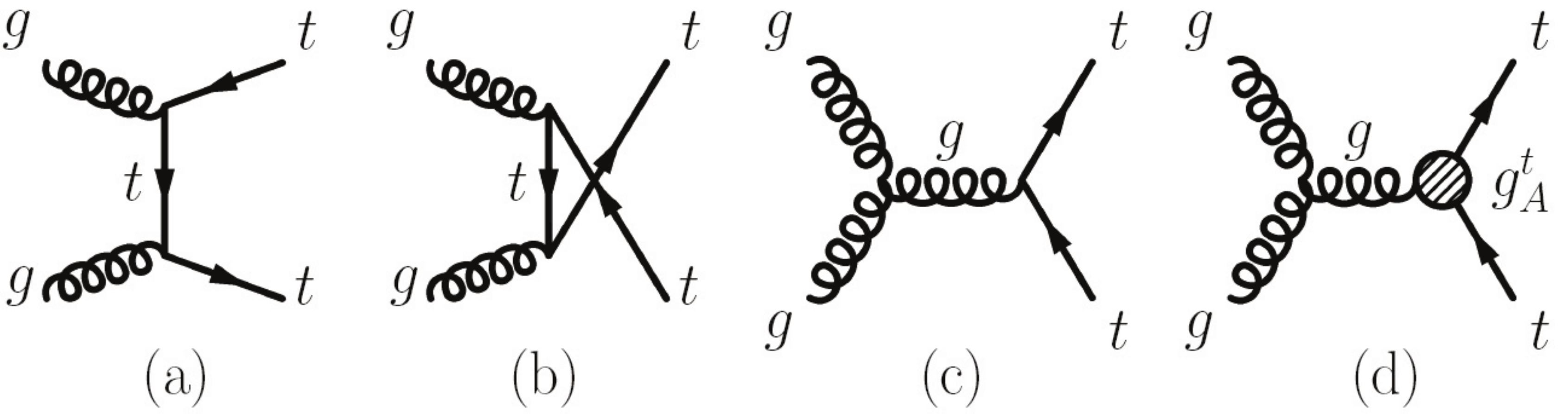}
\vspace{-0.3cm}
\caption{Feynman diagrams (a)-(d)
for the $g g \to t \bar t$ process, with
the contribution of the gluon effective axial-vector coupling to
the top-quark $g^{t}_A$.}
\label{fig:ggtt}
\end{center}
\end{figure}

The main contribution at the LHC to the top antitop-quark production, is given
by the gluon-gluon fusion process
\bea
g(p_1) g(p_2)\to t(p_3)\bar t(p_4)\, .
\label{ggprocess}
\eea
The Feynman diagrams (a)-(d) relative to
$gg\to t\bar t$, including the gluon axial-vector coupling,
are shown in Fig. \ref{fig:ggtt}

The polarized total cross sections
in the helicity basis and in the ZMF are given by
\bea
\sigma^{gg}_{LL}(\hat{s})&=&\frac{ \pi \alpha_S^2}{192\, \hat{s} \beta}
\left\{2\left(16-14\rho+31\rho^2\right)
-\frac{\rho}{\beta}\left(2+\rho\left(29+2\rho\right)\right)
\log{\frac{1+\beta}{1-\beta}}\right\},
\nonumber\\
\sigma^{gg}_{LR}(\hat{s})&=&\frac{ \pi \alpha_S^2}{192\, \hat{s}\beta}
\left\{2\left(11\left(\rho-4\right)+6|g_A^t|^2
\left(1-\rho\right)^2\right)+\frac{1}{\beta}
\left(32+(2-\rho)\rho\right)\log{\frac{1+\beta}{
1-\beta}}
\right\},
\nonumber\\
\sigma^{gg }_{RR}(\hat{s})&=&\sigma^{gg}_{LL}(\hat{s}),
\nonumber\\
\sigma^{gg}_{RL}(\hat{s})&=&\sigma^{gg}_{LR}(\hat{s}),
\label{crosspolgg}
\eea
where the symbols $\beta$ and $\rho$ are the same as defined above.
We have explicitly checked that the results in 
Eqs.(\ref{crosspolqq}) and (\ref{crosspolgg}) are separately 
gauge invariant for each $t\bar{t}$ polarizations, 
including the contribution from the gluon axial-vector coupling.
The sum over the  $t\bar{t}$ polarizations
reproduces the results for the unpolarized total cross section 
\cite{Gabrielli:2011zw}.
In appendix we report the corresponding expressions 
for the amplitude of $gg\to t\bar{t}$ process in the helicity
basis in the ZMF, and their square moduli given for all possible final spin 
configurations.

As we can see from Eq.(\ref{crosspolgg}), 
 the $gg\to t\bar t$ process turns out to be  
symmetric under the LR symmetry, even including the effect of the 
axial-vector coupling. This because of the C-parity of the initial 
gluon-gluon state. Therefore, the gluon axial-vector contribution to the 
LR asymmetry purely originates from the quark -antiquark fusion process. 
Then, the LR polarization asymmetry 
is very sensitive to the $q\bar{q}$ production mechanism, as in the 
case of the FB asymmetry. However, in the SM the FB or charge asymmetry
gets the leading contribution from a quantum interference effect in QCD
\cite{Kuhn:1998kw,Kuhn:1998jr,Bowen:2005ap,Kuhn:2011ri},
while the LR polarization asymmetry mainly comes from the 
interference of the tree-level QCD amplitude with the weakly corrected one. 
Therefore, at the LHC energies, the SM 
LR polarization asymmetry turns out to be at the level of few permille, 
while the FB asymmetry can be larger and at the level of few percent.
Then, due to the suppressed SM contribution,
the LR polarization asymmetry turns out to be 
a more sensitive probe of the NP scale 
$\Lambda$ associated to the gluon axial-vector form factor, with respect to 
the charge or FB asymmetry.

Finally, the corresponding hadronic cross sections  $pp\to t \bar t X$ at LHC 
for the polarized processes are 
obtained by convoluting the polarized partonic cross sections 
$\sigma_{qq}^{AB}$, $\sigma_{gg}^{AB}$, 
in Eqs.
(\ref{crosspolqq}),(\ref{crosspolgg}) respectively (where $A,B$ generically
indicate the L,R  polarization states of $t\bar{t}$ ), 
with the corresponding parton
distribution functions (PDF) for quarks and gluons, namely 
\bea
\sigma^{AB}_{p p \to t\bar t X}={\int \left( \sum_qd \rho_q
\sigma^{AB}_{qq}(\hat s)+ d\rho_{g}\sigma^{AB}_{gg}(\hat s)\right)},
\label{xsec} 
\eea 
where $d \rho_{q}$ and $d \rho_{g}$ indicate the
differential integrations in $dx_1 dx_2$ convoluted with the
quarks and gluon PDF, respectively. In the numerical integration of
Eq.~(\ref{xsec}) we have used the CTEQ6L1 parton distribution
function (PDF) \cite{Pumplin:2002vw}, where we set the PDF scale $\mu$
and the strong coupling constant $\alpha_S(\mu)$
at the same scale $\mu=m_t$, with top-quark mass $m_t=172$ GeV.

\section{Numerical results}

\subsection{Spin-correlation}
Recently ATLAS \cite{ATLAS:2012ao} and CMS \cite{CMS:2102}
collaborations have reported the 
measurements of the spin correlations
in $t\bar{t}$ production at the LHC.
The degree of correlation $A$ of $t\bar{t}$
system is defined as the fractional difference between the number of events
where the top and antitop quark spin orientations are aligned and those
where the top quark spins have opposite alignments, namely
\bea
A=\frac{N(\uparrow \uparrow) +N(\downarrow \downarrow)-
N(\uparrow \downarrow) -N(\downarrow \uparrow)}
{N(\uparrow \uparrow) +N(\downarrow \downarrow)+
N(\uparrow \downarrow) +N(\downarrow \uparrow)},
\label{Ah}
\eea
where the arrows denote the spins of the top and antitop 
with respect to a chosen quantization axis.
In the following we will indicate with $A_h$ the spin correlation $A$
evaluated in the helicity basis and in the ZMF of the $t\bar{t}$ pair.

The ATLAS collaboration has reported 
the following measurement for $A$ in the helicity basis ($A_h$)
\cite{ATLAS:2012ao}
\bea
A^{\rm ATLAS}_h=0.40^{+0.09}_{-0.08}\, ,
\label{ATLAS}
\eea
corresponding to an integrated luminosity of 2.1${\rm fb}^{-1}$. Candidate events
were selected in the dilepton topology with large missing transverse 
energy and at least two jets. The hypothesis of zero spin correlation is 
then excluded at 5.1 standard deviations.

On the other hand, the CMS collaboration, by using 5${\rm fb}^{-1}$ of 
integrated luminosity, has reported the following value for $A_h$ 
\cite{CMS:2102}
\bea
A^{\rm CMS}_h=0.24 \pm 0.02(\rm stat) \pm 0.08 (\rm syst)\, ,
\label{CMS}
\eea
where systematic and statistical errors are indicated in parenthesis.
The above results in Eqs.(\ref{ATLAS}),(\ref{CMS}) are inclusive 
in the available phase space of $m_{tt}$ invariant mass system. 

The corresponding SM prediction for LHC energies $\sqrt{S}=8$ TeV,
at the next-to-leading (NLO) order in QCD is \cite{Bernreuther:2010ny}
\bea
A^{\rm SM}_h=0.31 \, .
\label{SM}
\eea
The theoretical uncertainties, after including the NLO QCD corrections, 
due to the variation of 
factorization and renormalization scale, including the uncertainties on
parton distribution functions 
(PDF), are small and of the order of 1\% \cite{Bernreuther:2010ny}.
Although, the experimental central values in Eqs.(\ref{ATLAS}) and (\ref{CMS})
are quite different, the two measurements are compatible with each other and
with the SM prediction within 2 standard deviations. 

At this point, one may wonder if the above ATLAS and CMS results
can provide enough information to constrain the present scenario 
in the critical range of $\Lambda \sim 1-1.3$ TeV, required for explaining
the Tevatron top-quark anomaly \cite{Gabrielli:2011jf}. 
Unfortunately, a direct comparison with these results
is not possible in the framework of the low energy approximation adopted 
in Eq.(\ref{gA}) with $F(q^2,\Lambda)$ constant, since the
 ATLAS and CMS measurements in (\ref{ATLAS}) and (\ref{CMS}) are 
inclusive in the $m_{tt}$ invariant mass. Indeed, 
unitarity and perturbation theory restrict 
the validity of this approach
to the kinematic regions $m_{tt} < \Lambda$. 
In order to extend our predictions to 
the higher $m_{tt}$ invariant masses 
$m_{tt} > \Lambda$ we need to provide a shape for the form factor
$g_A(q^2)$ as a function of $q^2$. The price to pay  would be in this case
the introduction of new free parameters. A simple choice is to assume
a particular shape for the $g_A(q^2)$ function that tends to some
fixed cut-off in the regions $|q^2|=m^2_{tt}\gg \Lambda^2$, 
while reproducing 
the low energy limit of QCD Ward identities in Eq.(\ref{ginv}).
The purpose of this analysis is to determine the sensitivity of the
 inclusive top-spin correlation observables to this cut-off, at fixed values of the scale $\Lambda$.
We will present a detailed 
discussion on this issue in the last subsection. Now, we will
focus on the numerical analysis of the spin-correlation and LR asymmetries, 
in the low enery limit, that is
when we restrict our analysis to the regions $m_{tt} < \Lambda$.

Following the low-energy approach of Refs.\cite{Gabrielli:2011jf},\cite{Gabrielli:2011zw}, 
in order to simplify the analysis we will assume a real and 
universal gluon axial-vector coupling, and reabsorb all the NP effects 
in the scale $\Lambda$ defined as follows
\bea
g^{(q,t)}_A(q^2)&=&\frac{q^2}{\Lambda^2}\, .
\label{gAEFT}
\eea
where we neglected any potential logarithm 
contribution proportional to $q^2\log(q^2/\Lambda^2)$ and 
higher powers of $q^2/\Lambda^2$ terms. This has the advantage of performing
a phenomenological model independent analysis, by introducing only one 
relevant free parameter.
The quark universality of the gluon axial-vector coupling is not only a
reduction of the free parameters of the model, but it is actually supported by
the explanation of the Tevatron top-quark asymmetry anomaly 
in terms of this scenario \cite{Gabrielli:2011jf}. Therefore, from now on,
we will omit from our notations the quark flavor $q$ dependence in the gluon 
axial-vector coupling.

\begin{figure}[t]
\begin{center}
\includegraphics[angle=-90, width=0.45 \textwidth]{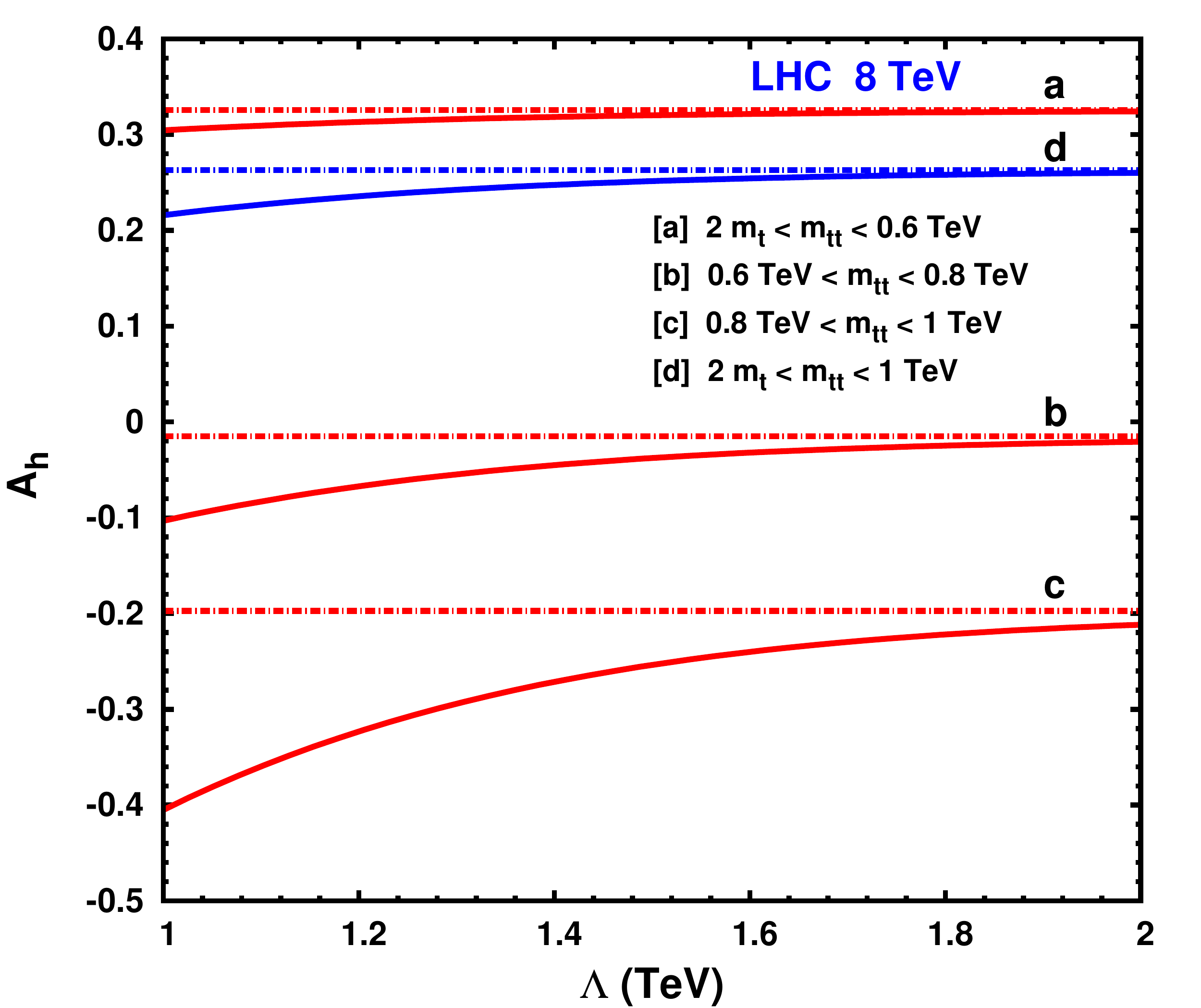}
\includegraphics[angle=-90, width=0.45 \textwidth]{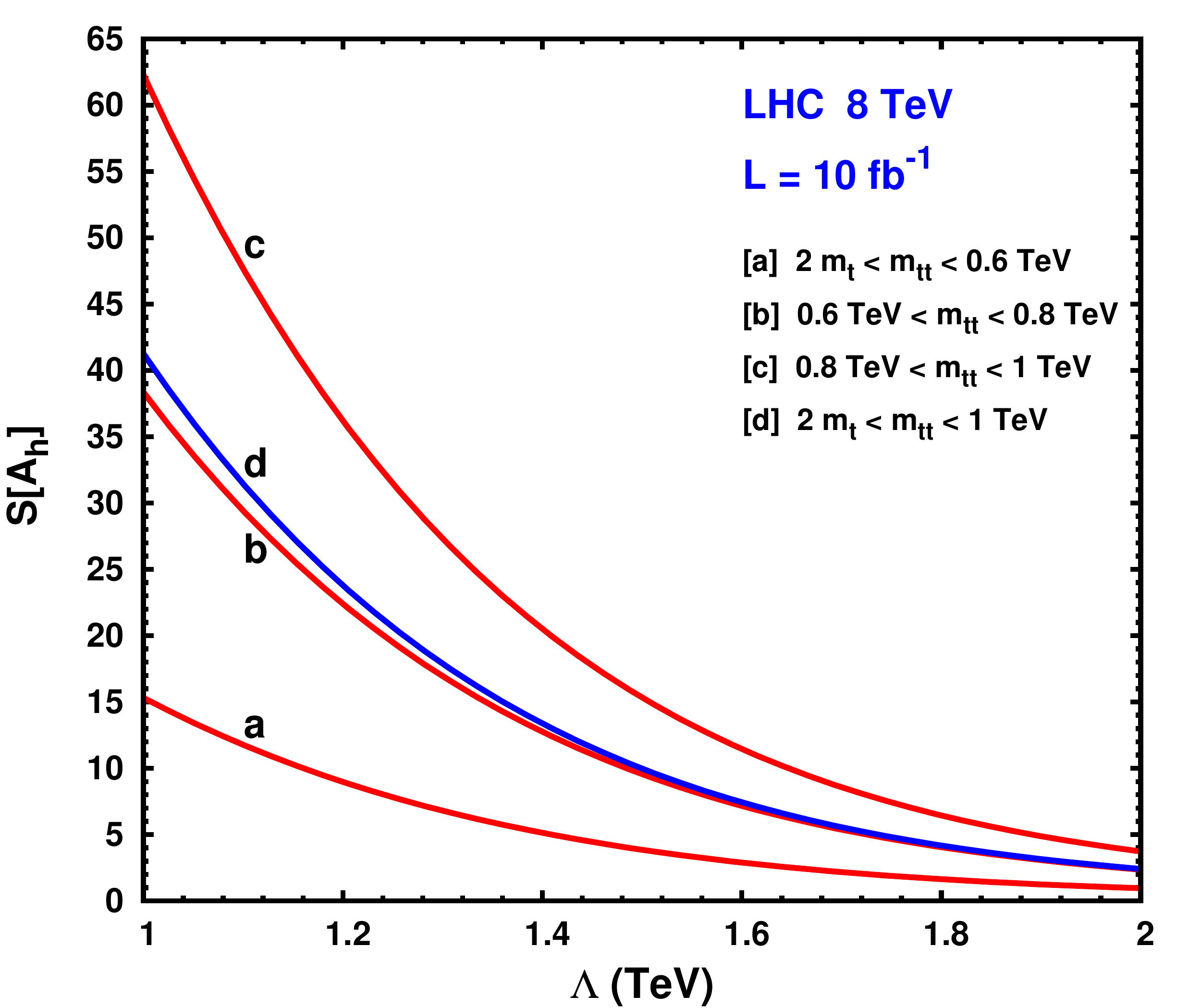}
\vspace{-0.3cm}
\caption{Top-antitop spin correlation 
$A_h$ (left plot) and corresponding significance 
$S[A_h]$
(right plot) in $t\bar{t}$ events, in the helicity basis and
laboratory frame, for the LHC center of mass energy of 8 TeV and
integrated luminosity L=10 fb$^{-1}$, as a 
function of the scale $\Lambda$ and for some ranges of $m_{tt}$. Dashed
lines in the left plot correspond to the SM predictions at leading order 
in QCD.}
\label{fig3}
\end{center}
\end{figure}

In Figs.\ref{fig3} 
we present our numerical results for the 
spin-correlation observable $A_h$ in Eq.(\ref{Ah}) (left plot) and
its corresponding statistical significance  $S[A_h]$ (right plot), evaluated
for LHC energies of $\sqrt{S}=8$ TeV,
in the helicity basis, and in the laboratory frame. 
We show our results for 
some kinematic ranges of $m_{tt}$ and $\Lambda$ in the range 
$1\,  {\rm TeV}<\Lambda<2\, {\rm TeV}$. The kinematic ranges of $m_{tt}$ 
considered in our analysis are the following
\bea
[{\rm a}]&=& 2\, m_t < m_{tt} < 0.6\, {\rm TeV},\\ \nonumber
[{\rm b}]&=& 0.6\, {\rm TeV}  < m_{tt} < 0.8\, {\rm TeV},\\ \nonumber
[{\rm c}]&=& 0.8\, {\rm TeV} < m_{tt} < 1\, {\rm TeV},\\ \nonumber
[{\rm d}]&=& 2\, m_t < m_{tt} < 1\, {\rm TeV} \, .
\label{rangemtt}
\eea
The dashed lines correspond to 
the SM prediction at the LO in QCD. As we can see from the left plot in 
\ref{fig3}, 
the spin-correlation is positive for the [a] and [d] ranges, while it changes
sign for the [b] and [c] range. Although we choose the convention 
Re$[g_A]>0$, the spin-correlation $A_h$  and cross sections 
do not depend on the sign[$g_A$].

As we can see from the results in Fig.\ref{fig3},  the common trend
of this scenario is a decrease of $A_h$ with respect to the SM prediction,
while the corresponding SM deviations increase by selecting 
kinematic regions of $m_{tt}$ masses 
close to the scale $\Lambda$. This last property is due to the fact that
the axial-vector form factor $g_A$ grows quadratically with  $m_{tt}$.
On the other hand, the common decrease from the SM prediction
can be easily understood by looking at the definition of $A$ in Eq.(\ref{Ah}) 
and at the polarized cross sections in Eqs.(\ref{crosspolqq}),
(\ref{crosspolgg}).
The gluon-gluon fusion mechanism dominates at the LHC energies
with respect to the quark-antiquark annihilation process. 
In this case, $|g_A|$ enters only through
the combination $\sigma^{gg}_{LR}+\sigma^{gg}_{RL}$ combination,
since   $\sigma^{gg}_{LL}$ and $\sigma^{gg}_{RR}$ do not depend on $g_A$.
This results in a positive contribution to the total cross section 
(cfr. Eqs.(\ref{crosspolqq}) and (\ref{crosspolgg})), but a
negative one in the numerator of $A_h$, see  Eq.(\ref{Ah}), giving rise 
to a destructive contribution with respect to the SM one.

\begin{figure}[t]
\begin{center}
\includegraphics[angle=-90, width=0.45 \textwidth]{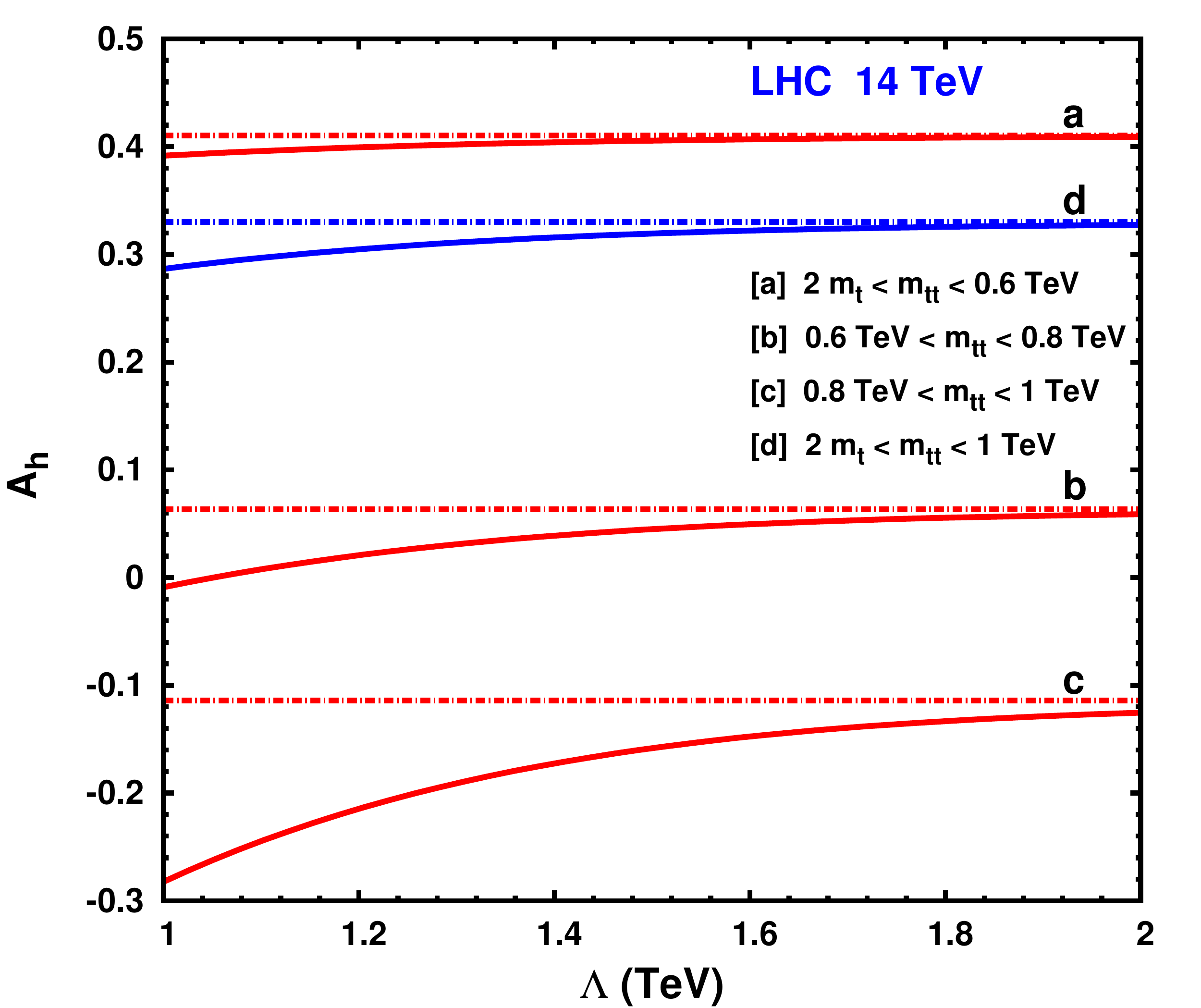}
\includegraphics[angle=-90, width=0.45 \textwidth]{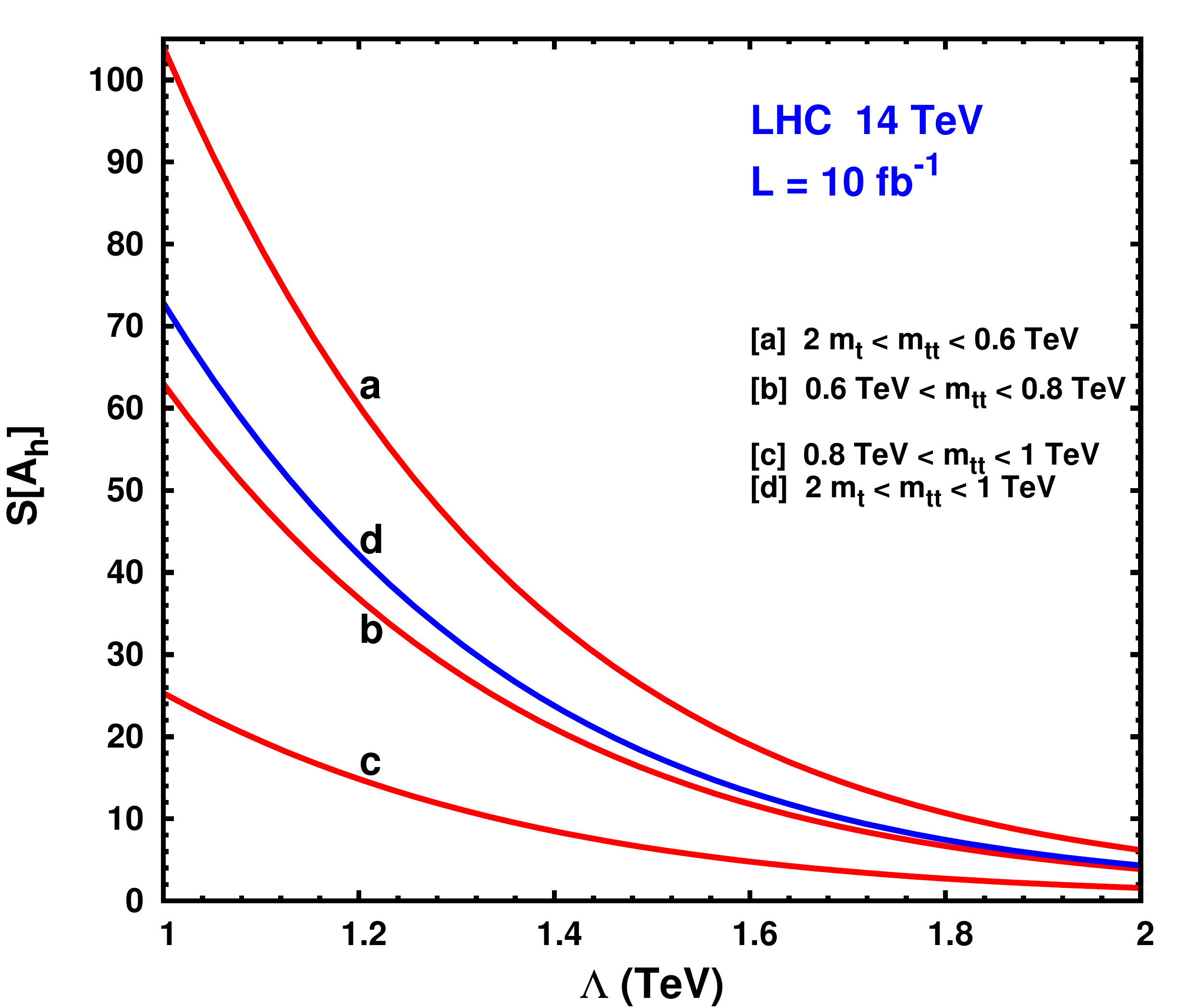}
\vspace{-0.3cm}
\caption{Same as in Fig. \ref{fig3} but for the LHC center of mass energy 
of 14 TeV  and integrated luminosity L=10 fb$^{-1}$.}
\label{fig4}
\end{center}
\end{figure}

For the [a] and [d] ranges in the left plot of Fig.\ref{fig4}, 
the maximum deviation from the SM value
is obtained for  $\Lambda = 1$ TeV, corresponding to a 10\% deviation
from the SM prediction, 
while for the [b] and [c] ranges the effect is larger reaching almost 
a 25\% and 100\% deviations for the [c] and [d] ranges respectively. 
For values of 
$\Lambda =2$ TeV the overall NP effect is strongly reduced 
and $A_h$ results are much closer to the 
corresponding SM ones. The numerical values of $A_h$ in Fig.\ref{fig3} for
$\Lambda=1$ TeV and $\Lambda=2$ TeV are 
$A_{h}=(30,-10,-40,22)$\% and $A_{h}=(32,-2.1,-21,26)$\% respectively. 
The four series of numbers reported in parenthesis will indicate 
from now on, if not differently specified,
the results corresponding to the 
$m_{tt}$ integration ranges of [a],[b],[c],[d] respectively.

On the right plot of Fig.\ref{fig3}, 
we show the corresponding statistical significance for $A_h$, that, following 
the definition of spin-correlation $A$ in (\ref{Ah}), is
\bea
S[A_h]&=&\Delta A_h \sqrt{\sigma^{\rm SM+NP} L}\, ,
\label{sign}
\eea
where $\Delta A_h = |A_h^{\rm SM+NP}-A_h^{\rm SM}|$,  $\sigma^{\rm SM+NP}$ is the total unpolarized cross section, and 
$L$ stands for the integrated luminosity, 
while the $\scriptstyle{\rm SM+NP}$ suffix stands 
for the full SM and NP contribution.
In $\sigma^{\rm SM+NP}$ we have used 
the LO QCD cross sections. 
Notice  that the significance in Eq.(\ref{sign})
is a simple theoretical estimation of the true one, since
it does not take into account detector efficiencies,
acceptance, resolution and systematics.
From the results in the right plots of Fig.\ref{fig3}, we can see that the 
corresponding significances for $L=10\, {\rm fb}^{-1}$ are quite large. In
particular, for $\Lambda=1$ TeV and $\Lambda=2$ TeV we get  
 $S[A_{h}]=(15,38,62,41)$ and $S[A_{h}]=(0.9,2.3,3.7,2.4)$ respectively.
We can see that, for $\Lambda=2$ TeV, the significance is considerably lower, 
with the maximum effect $S[A_h]\simeq 4$ corresponding to the range [c]. 
Therefore, we stress that by analyzing the $m_{tt}$ distributions of
$t\bar{t}$ spin correlations at the LHC, 
the full range up to $\Lambda \sim 2$ TeV can be probed at LHC 8 TeV, even 
with an integrated luminosity of $L=10\, {\rm fb}^{-1}$.

In Fig.\ref{fig4}  we present the corresponding results of
Fig.\ref{fig3}, but for LHC energies
of $\sqrt{S}=14$ TeV and integrated luminosity $L=10\, {\rm fb}^{-1}$. 
By increasing the LHC center of mass energy, we see that $|A_h|$ increases
by roughly 25-30\% with respect to the corresponding values at 
$\sqrt{S}=8$ TeV in the regions $A_h>0$, while it decreases of 
roughly the same amount in the regions $A_h<0$,
for almost all the $m_{tt}$ ranges [a-d], including the SM values. 
In particular, we get $A_{h}=(39,-0.9,-28,28)\%$ 
and  $A_{h}=(40,5.9,-12,33)\%$  for $\Lambda=1$ and $\Lambda=2$ TeV 
respectively, where the latter are quite close to the SM values.
Due to the larger cross sections, 
the corresponding significances, with respect to the corresponding results at
$\sqrt{S}=8$ TeV , are also increased,  
roughly by 70\% and 30\% effects for
$\Lambda=1$ TeV and $\Lambda=2$ TeV respectively. In particular,  
we get $S[A_{h}]=(25,63,104,73)$ and $S[A_{h}]=(1.5,3.8,6.1,4.3)$ 
for $\Lambda=1$ TeV and $\Lambda=2$ TeV respectively.

\subsection{ Left-Right spin asymmetry}
Here we consider the LR polarization asymmetry $A_{LR}$ defined as
\cite{Kao:1999kj}
\bea
A_{LR}&=&\frac{N(\uparrow \uparrow) -N(\downarrow \downarrow)+
N(\uparrow \downarrow) -N(\downarrow \uparrow)}
{N(\uparrow \uparrow) +N(\downarrow \downarrow)+
N(\uparrow \downarrow) +N(\downarrow \uparrow)},
\label{ALR}
\eea
where the left and right arrows denote the spins of the top and antitop 
respectively, with respect to a chosen quantization axis.
As mentioned in the introduction, 
the SM contribution to this asymmetry ($A_{LR}^{\rm SM}$) 
is suppressed, being induced by one loop weak radiative corrections to the QCD 
$q\bar{q}\to t\bar{t}$ production. The typical SM value for $A_{LR}^{\rm SM}$
is very small, being
of  order of 0.5\% and 0.04\% for the cases of LHC 14 TeV and  Tevatron
respectively \cite{Kao:1999kj}. Therefore, this is a very sensitive 
probe to any potential parity-violating new physics beyond the SM. 
The LR polarization asymmetry has been 
analyzed in \cite{Li:1997ae,Li:1997gh,Kao:1994rn,Sullivan:1996ry} for the Tevatron and in \cite{Kao:1999kj,Cao:2010nw} for LHC, mainly in the framework 
of minimal supersymmetric extensions of the SM \cite{Kao:1999kj,Li:1997ae,Li:1997gh,Sullivan:1996ry} and more recently in more exotic NP scenarios like 
axigluons, third-generation enhanced LR models, and supersymmetric models
without R-parity \cite{Cao:2010nw}.

We will see that in our framework, the $A_{LR}$ 
is at least one order of magnitude larger than 
the corresponding SM contribution, since it is
induced at the tree-level by the effect of the 
axial-vector coupling of the gluon. For this reason 
we will neglect the SM  contribution to $A_{LR}$ in our analysis.
Accordingly, we will use the following formula 
for the corresponding significance $S[A_{LR}]$
\bea
S[A_{LR}]&=&|A^{NP}_{LR}| \sqrt{\sigma^{\rm NP+SM} L}\, ,
\label{sign2}
\eea
where in $A^{NP}_{LR}$ the leading contribution to the asymmetry is 
induced by the Re$[g_A]$ terms, which appears in the numerator 
of the right hand side of Eq.(\ref{ALR}), while the 
denominator clearly includes the NP and SM contributions. 

\begin{figure}[t]
\begin{center}
\includegraphics[angle=-90, width=0.45 \textwidth]{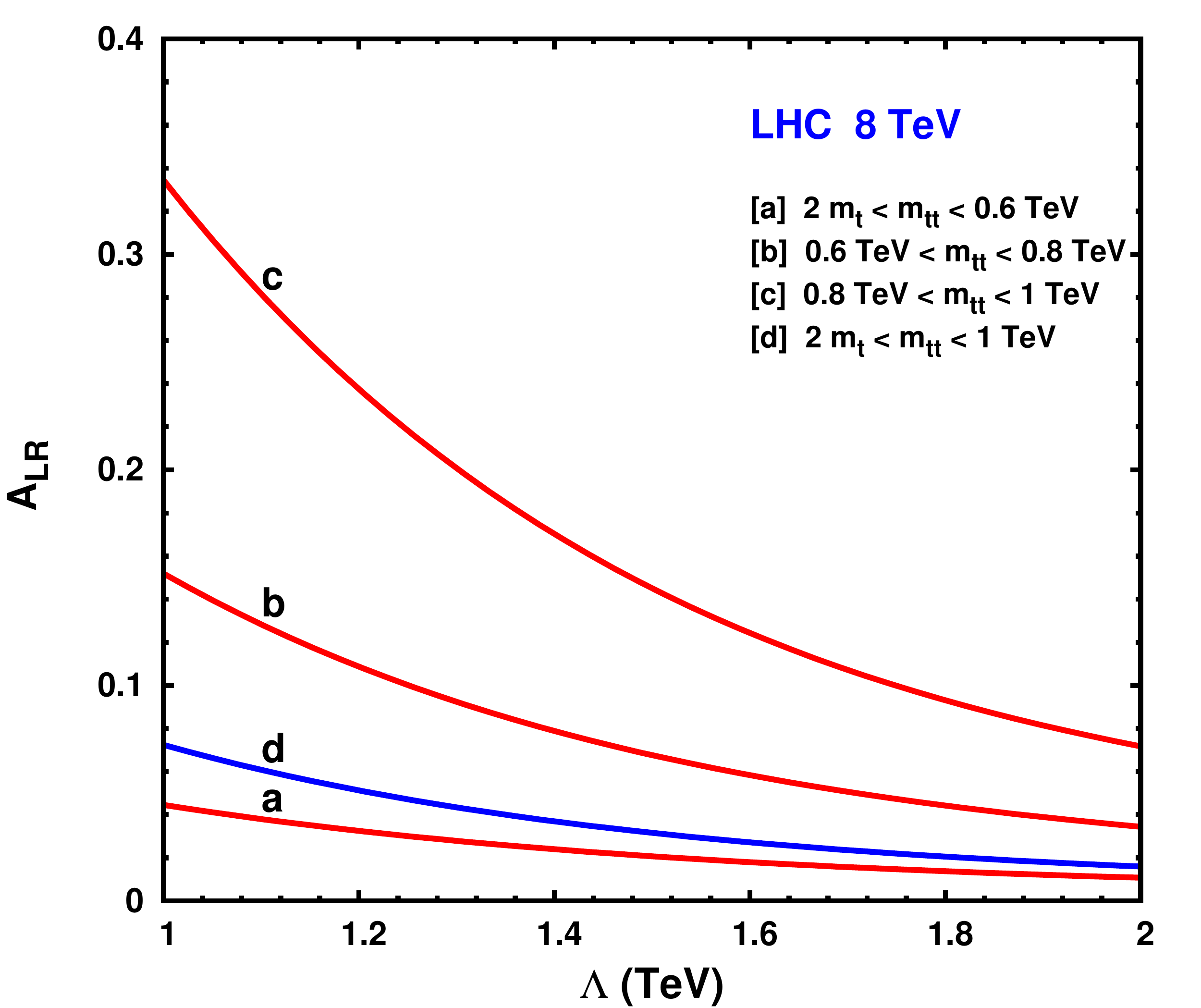}
\includegraphics[angle=-90, width=0.45 \textwidth]{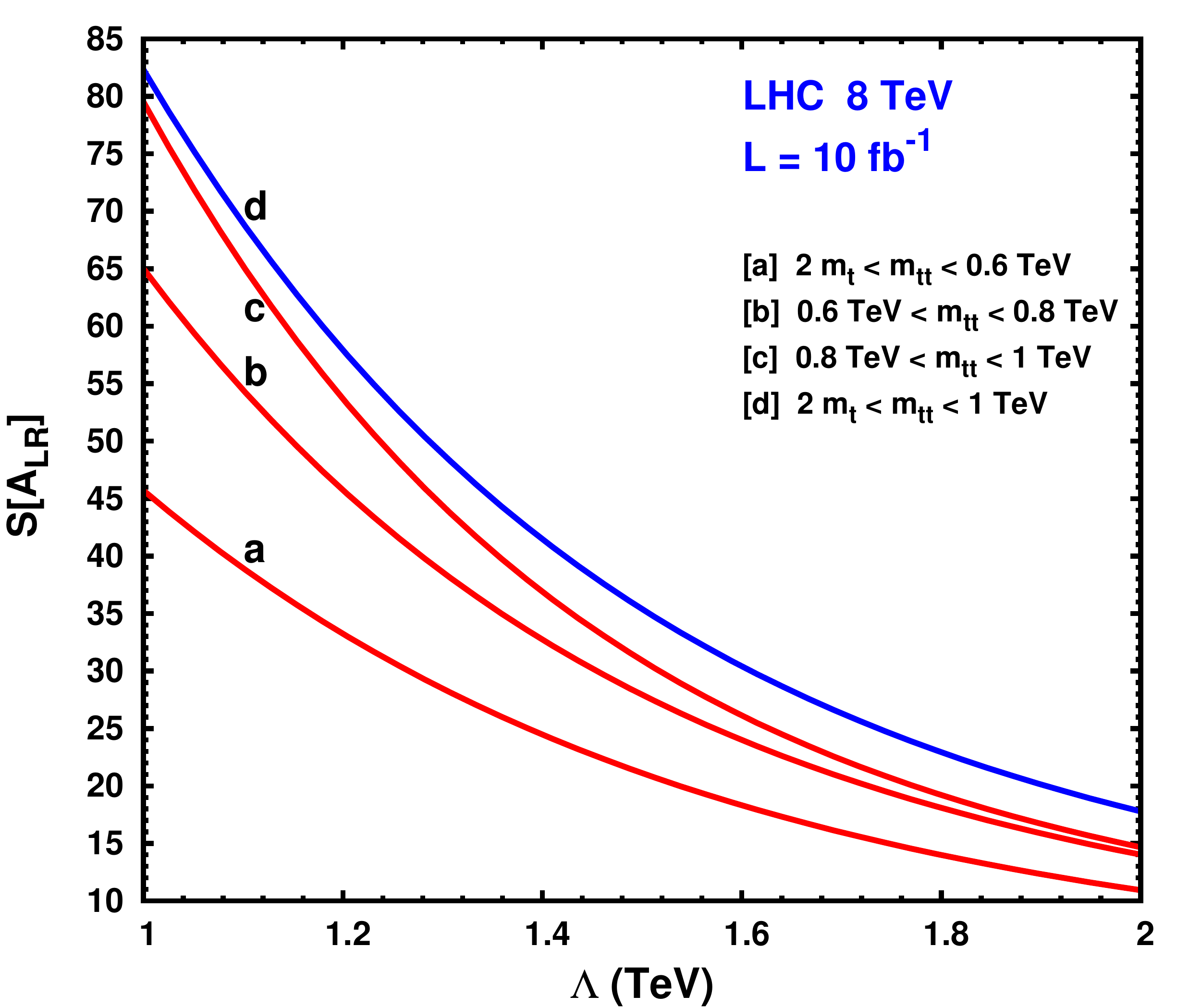}
\vspace{-0.3cm}
\caption{Top-antitop left-right polarization asymmetry  $A_{LR}$ (left plot) 
and corresponding significance $S[A_{LR}]$ (right plot) in
$t\bar{t}$ events, in the helicity basis and
laboratory frame, for the LHC center of mass energy of 8 TeV and
integrated luminosity L=10 fb$^{-1}$, as a 
function of the scale $\Lambda$ and for some ranges of $m_{tt}$.}
\label{fig5}
\end{center}
\end{figure}

In the left plot of Fig.\ref{fig5} we present our results 
for the $A_{LR}$  calculated 
at the LO in QCD and in the helicity basis and laboratory frame, while
on the right plot we show the corresponding significance $S[A_{LR}]$ 
for $L=10\, {\rm fb}^{-1}$.
From these results we can see that the contribution induced by the 
pure axial-vector coupling to $A_{LR}$ is sizeable. In particular, 
for  $\Lambda=1$ TeV, we get $A_{LR}=(4.5,15,33,7.2)$\%, with a corresponding 
significance $S[A_{LR}]=(45,65,79,82)$, 
while for $\Lambda=2$ TeV the value of $A_{LR}$ lowers considerably, namely 
$A_{LR}=(1.1,3.4,7.2,1.6)$\% with a corresponding 
significance $S[A_{LR}]=(11,14,15,18)$.
From these results we can see that, although $A_{LR}$ is smaller than 
$A_h$, its statistical 
significance
is higher than the corresponding one of $A_h$, 
mainly due to the fact that in the $A_{LR}$ the SM background 
is negligible. Therefore, $A_{LR}$ is a more sensitive probe of this scenario 
than $A_{h}$.

Notice that the sign of $A_{LR}$ in the right plot of 
Fig.\ref{fig5} depends on the convention we used for 
the sign of ${\rm Re}[g_A]$, namely positive. 
If we switch this sign, the asymmetry change
sign too, being directly proportional to  ${\rm Re}[g_A]$.
Therefore, we stress that a non vanishing measurement of $A_{LR}$,
also determines the sign of ${\rm Re}[g_A]$ in the framework of 
this scenario.

\begin{figure}[t]
\begin{center}
\includegraphics[angle=-90, width=0.45 \textwidth]{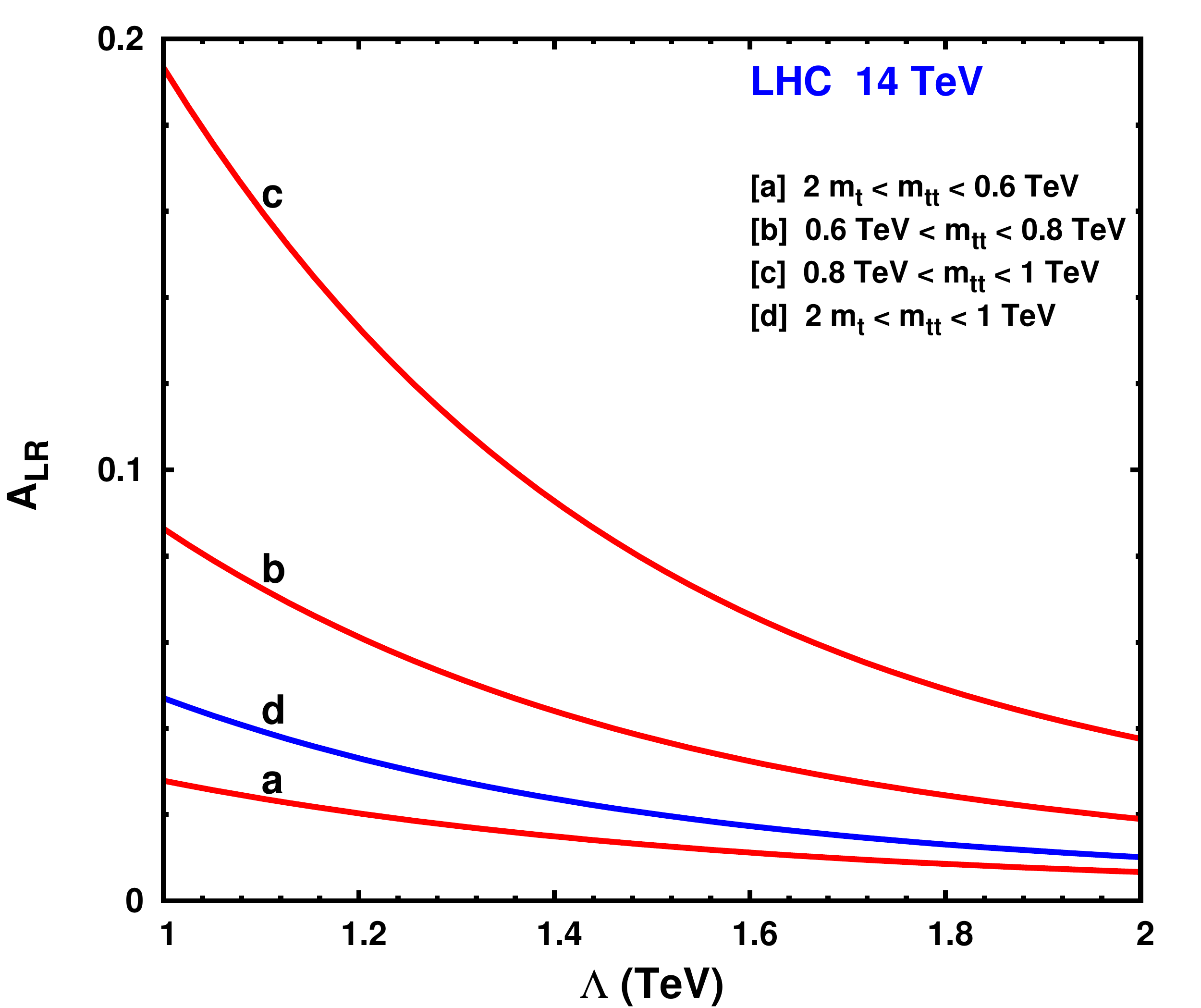}
\includegraphics[angle=-90, width=0.45 \textwidth]{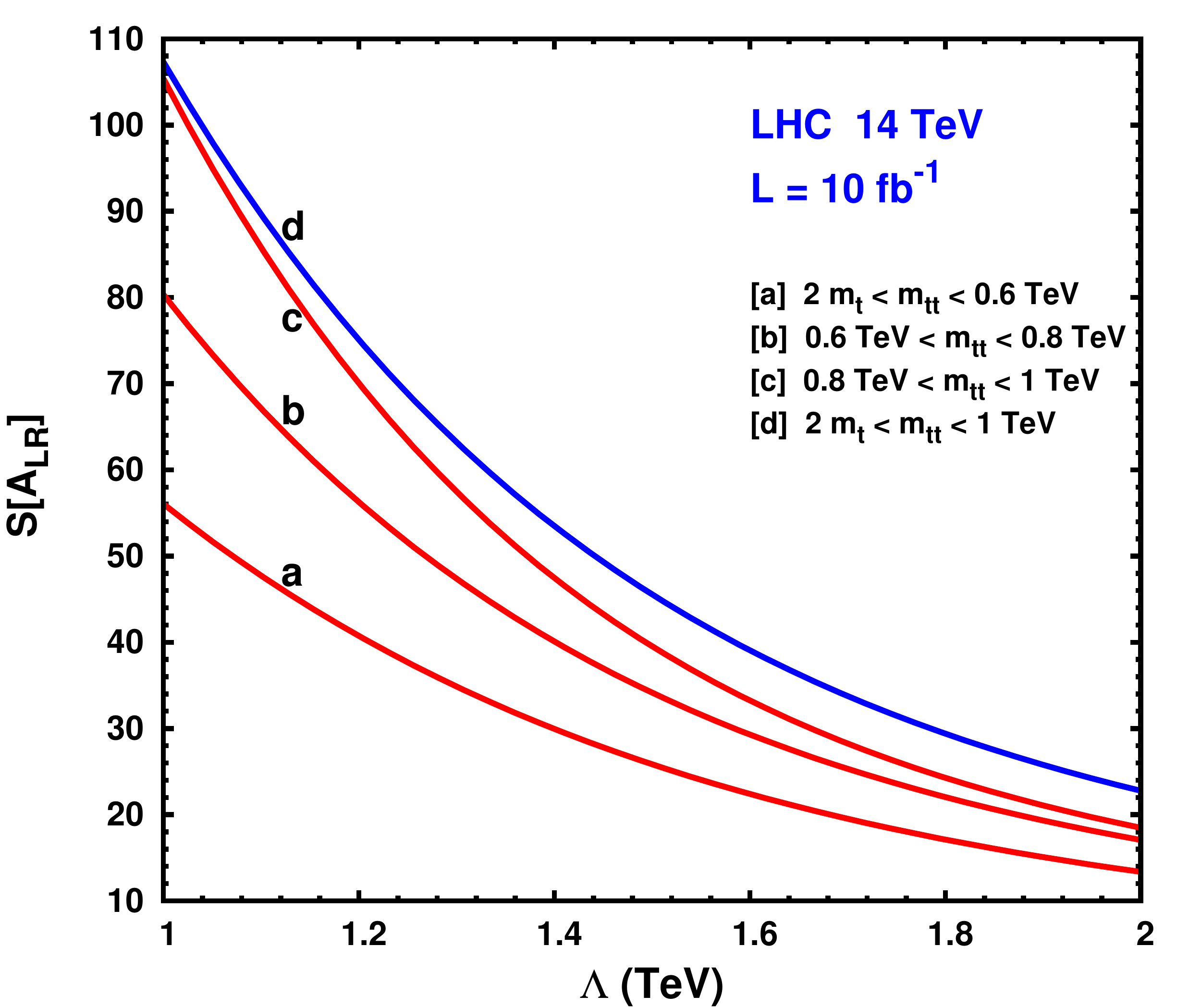}
\vspace{-0.3cm}
\caption{
Same as in Fig. \ref{fig4} but for LHC center of mass energy 
of 14 TeV   and integrated luminosity L=10 fb$^{-1}$. }
\label{fig6}
\end{center}
\end{figure}

In Fig. \ref{fig6} we show the corresponding results of $A_{LR}$ for LHC
energy of $\sqrt{S}=14$ TeV.
%%%%%%%%%%%%%%%%%%%%%
As we can see from the left plot of Fig.\ref{fig6}, the trend of $A_{LR}$ by 
increasing the LHC energy is different with respect to the corresponding $A_h$
behavior, at fixed values of $\Lambda$.
In particular, there is roughly a 45\% decrease in $A_{LR}$, 
when passing from $\sqrt{S}=8$ TeV to 14 TeV.
This is due to the fact that the total cross section, dominated by the 
gluon-gluon fusion process, grows faster than the
$\sigma^{qq}_{LR}-\sigma^{qq}_{RL}$ contribution, by 
increasing the center of mass energy.  In particular, for the [a-d] 
$m_{tt}$ ranges we get $A_{LR}=(2.8,8.6,19,4.7)\%$ 
and  $A_{LR}=(0.7,1.9,3.8,1)\%$ for $\Lambda=1$ TeV and $\Lambda=2$ TeV respectively.

On the other hand, by comparing the corresponding significances 
$S[A_{LR}]$ at $\sqrt{S}=8$ TeV and 14 TeV in the right plots of 
Figs.\ref{fig5} and \ref{fig6} respectively,
we see that there is an almost 30\% increase of $S[A_{LR}]$ in all integration 
ranges [a-d], when passing from $\sqrt{S}=8$ TeV  to 14 TeV.

\subsection{Comparison with ATLAS and CMS results}
Now we discuss the impact of this scenario 
on the $m_{tt}$ inclusive measurements of $A_h$ performed by ATLAS and CMS 
collaborations, corresponding to LHC data at $\sqrt{S}=$7 TeV.
In particular, we are interested in estimating the maximum effect
induced by the axial-vector
coupling  contribution to these inclusive observables at fixed values of 
$\Lambda$.
As mentioned before,  this can be done at the cost of 
introducing a new free parameter in addition to $\Lambda$, 
that should be understood as the upper bound on the 
$g_A$ form factor in the high $m_{tt}$ mass regions $m_{tt} \gg \Lambda$. 

Dimensional analysis and 
unitarity arguments, suggest that the $g_A(q^2)$ form factor should 
not grow with $|q^2|$ indefinitely and should tend
at most to a constant value in the asymptotic limit $|q^2|\gg \Lambda^2$, where
in our case this corresponds to $q^2=m_{tt}^2 \gg \Lambda^2$.
In order to implement this parametrization, we replace $g_A$
in Eq.(\ref{gA}) by some test function $g_A(q^2)=G_F(q^2)$ , 
that reproduces the low energy limit 
in Eq.(\ref{gA}), but satisfies the asymptotic condition 
$\lim_{|q^2|\to \infty}\{g_A(q^2)\}= \bar{g}_A$, where  $\bar{g}_A$ is some
constant. 
By naturalness arguments, we expect $\bar{g}_A$ to be at the most of order 
${\cal O}(1)$. For simplifying the analysis, we will restrict to the case 
in which the constant $\bar{g}_A$ is real.

Basically, $\bar{g}_A$ plays the role here 
of a new dimensionless free parameter that
parametrizes the upper bound of the axial-vector form factor 
$g_A$, in the kinematic regions $m_{tt} \gg \Lambda$. 
By using some test functions for the $g_A$ form factor, 
satisfying the above criteria, we will show that for $\Lambda > 1$ TeV, 
deviations from the SM results in the inclusive $A_h$ values
are very small, at most of order of 10\%  for asymptotic values of 
$\bar{g}_A \le 10 $. 

\begin{figure}[t]
\begin{center}
\includegraphics[angle=0, width=0.40 \textwidth]{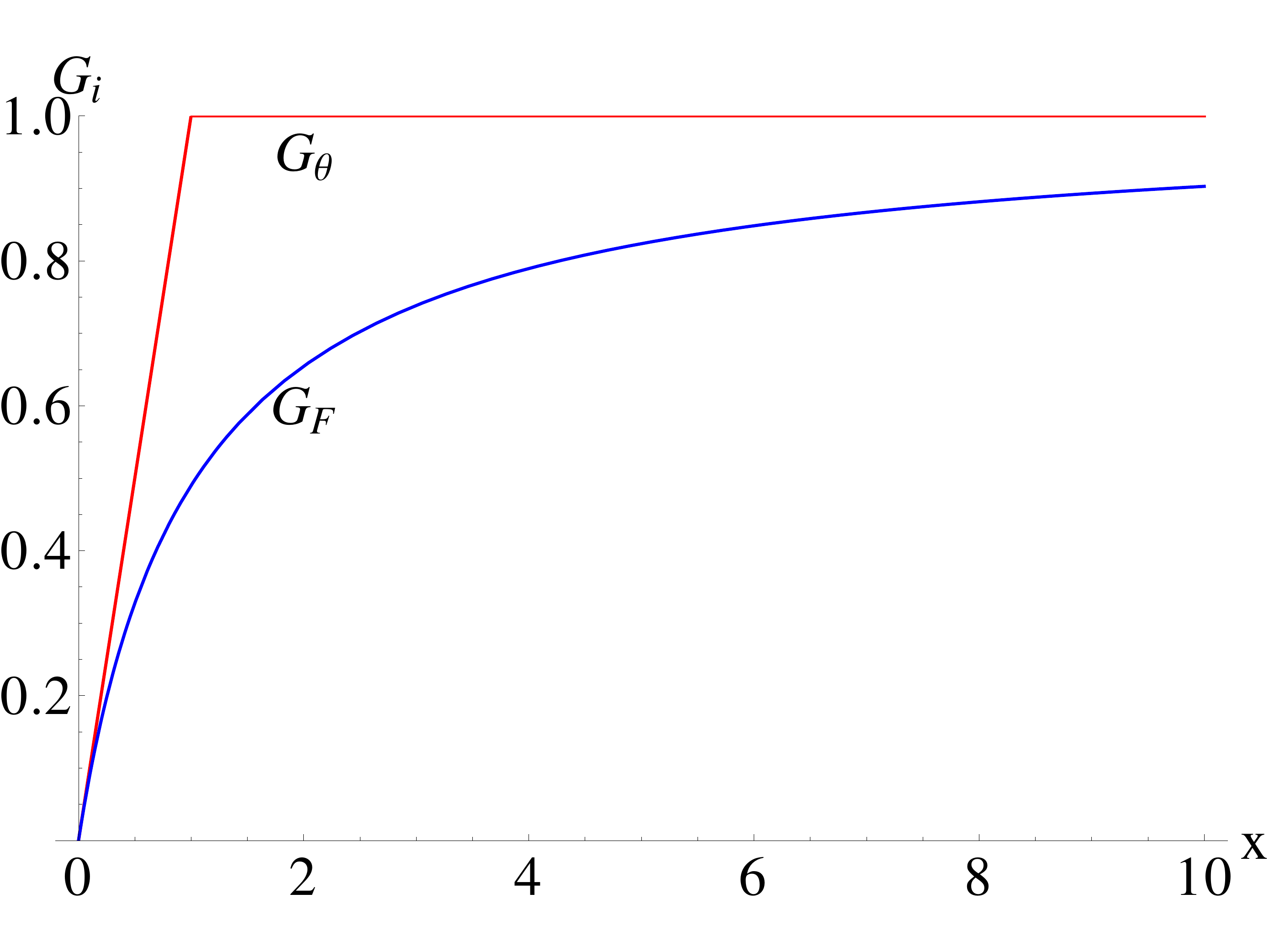}
\vspace{-0.3cm}
\caption{The $G_F(x)$, evaluated at $\bar{g}_A=1$, and 
$G_{\theta}(x)$ versus $x=\hat{s}/\Lambda^2$.} 
\label{fig7}
\end{center}
\end{figure}

As a toy model, we will use the following function
to parametrizes the form factor $g_A(q^2)=G_{F}(x)$, as a function of
$x=q^2/\Lambda^2$, namely
\bea
G_F(x)&=&\bar{g}_A -\log{\left(\frac{e^{\bar{g}_A}+y}{1+y}\right)},
{\rm with }~~~y\, =\, \frac{xe^{\bar{g}_A}}{(e^{\bar{g}_A}-1)}\, ,
\label{Gfunct}
\eea
where $G_{F}(x)$ 
satisfies the required conditions $G_{F}(x)=x+{\cal O}(x^2)$ for $x\ll 1$ and
$\lim_{x\to \infty} G_{F}(x)=\bar{g}_A$. In Fig.\ref{fig7}, we plot for 
comparison $G_{F}(x)$, evaluated at $\bar{g}_A=1$, with the function 
$G_{\theta}(x)$ defined as $G_{\theta}(x)=x \theta(1-x) + \bar{g}_A\theta(x-1)$.

\begin{figure}[t]
\begin{center}
\includegraphics[angle=-90, width=0.45 \textwidth]{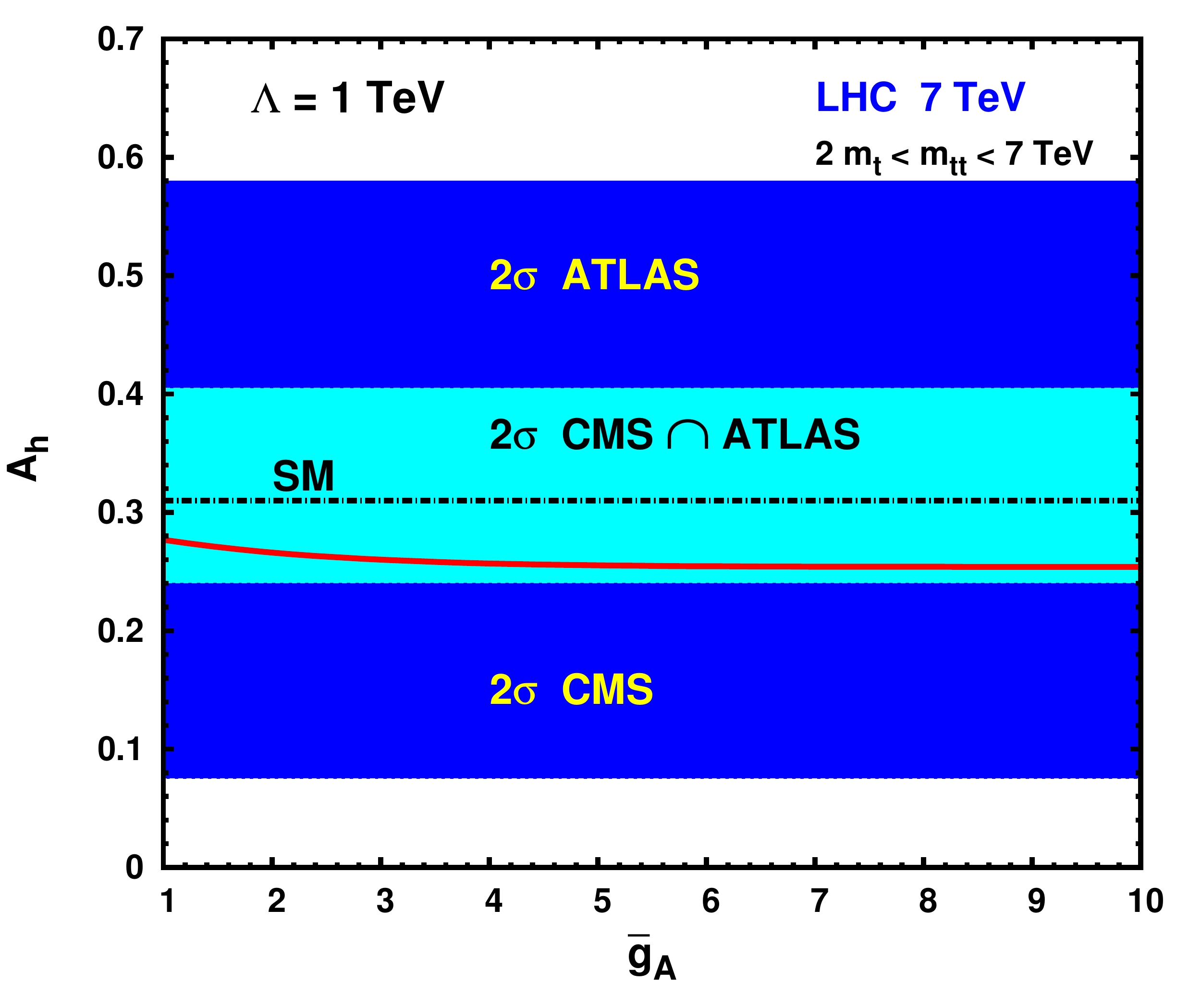}
\includegraphics[angle=-90, width=0.45 \textwidth]{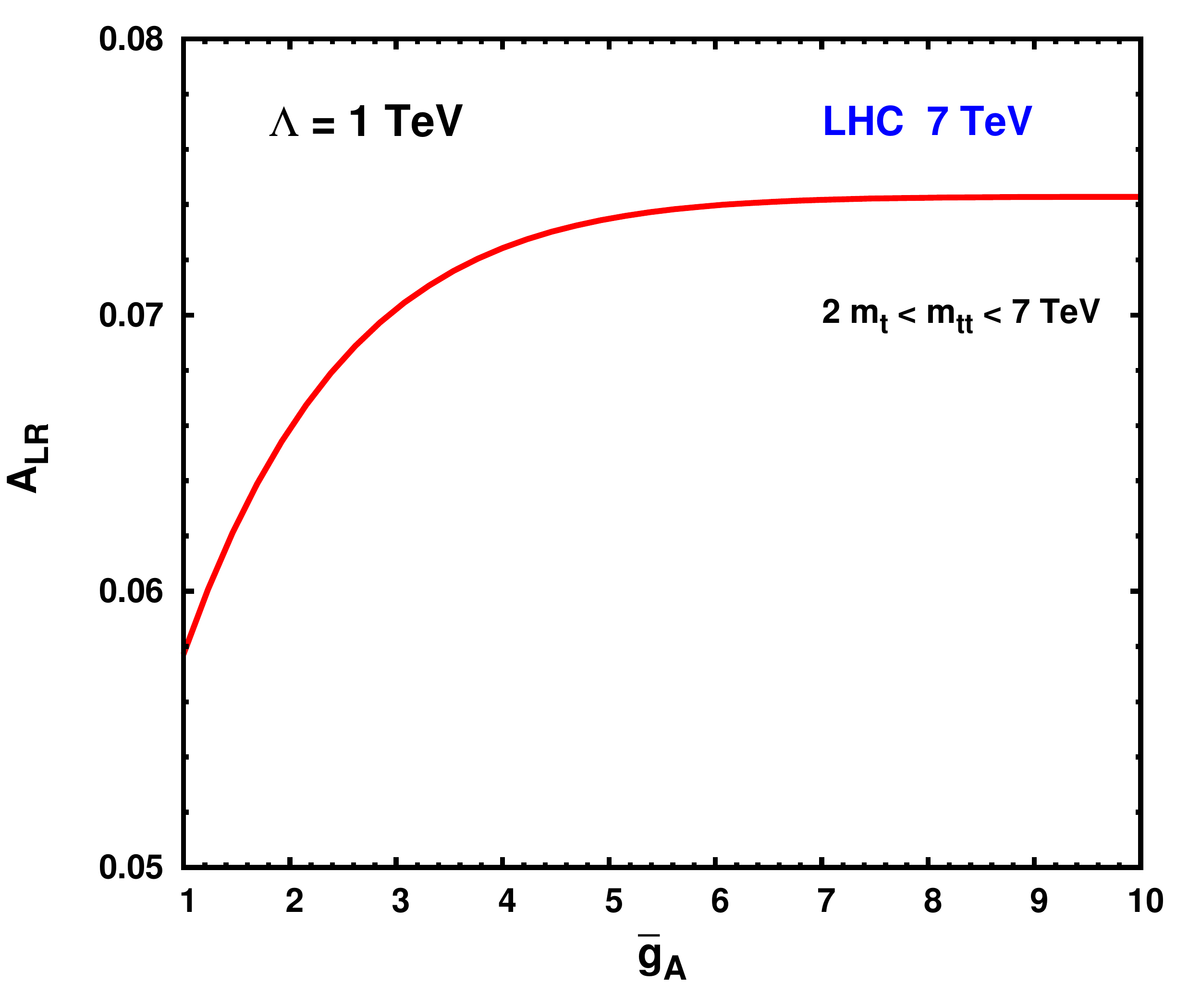}
\vspace{-0.3cm}
\caption{The $m_{tt}$ inclusive values for 
$A_h$ (left) and $A_{LR}$ (right) respectively, for the LHC 7 TeV energy and $\Lambda=1$ TeV, 
versus the gluon axial-vector coupling cut-off $\bar{g}_A$. In left plot, 
colored (dark blue) bands stand 
for the ATLAS (top region) 
and CMS (down region)  2$\sigma$ regions, while 
the middle (light blue) band is the overlap 
between these two areas. The dashed dot and continuous (red)
lines correspond
to the SM prediction in Eq.(\ref{SM}) at the NLO in QCD and 
to the prediction of our scenario for the $G_F$ function respectively,
all multiplied by the NLO rescaling factor.}
\label{fig8}
\end{center}
\end{figure}

In the left plot of Fig.\ref{fig8} we show our prediction
for the $m_{tt}$ inclusive spin correlation observable $A_h$ 
corresponding to $\Lambda=1$ TeV and   LHC energy of 7 TeV, as a function
of $\bar{g}_A$, for the $G_F(x)$ function. The colored bands stand for 
the 2$\sigma$ regions for the ATLAS (top region) \cite{ATLAS:2012ao} 
and CMS (down region) \cite{CMS:2102} measurements of $A_h$, while 
the middle band is the overlap 
between these two areas. The dashed dot and continuous (red) 
lines correspond
to the SM prediction in Eq.(\ref{SM}) at the NLO in QCD and 
to the prediction of our scenario for the $G_F$ function respectively,
suitably rescaled to the SM value at the NLO. In rescaling our predictions
we multiplied the results obtained at the LO in QCD by the SM $K$ factor
for the spin correlation defined as
$K=A^{\rm NLO}_h/A^{\rm LO}_h$ at $\sqrt{S}=7$ TeV.

As we can see from these results, the impact of this scenario
on the inclusive observable $A_h$ is a
decrease of the $A_h$ values with respect to the SM prediction.
In the region $4<\bar{g}_A < 10$,  the $A_h$ approaches to a plateau, namely
$A_h\sim 25\%$.  
The expected deviations from the SM prediction, for $\Lambda=1$ TeV,
are within the 2$\sigma$ bands of ATLAS and CMS measurements. 
If we consider larger values of the scale
$\Lambda > 1$ TeV, the SM deviations are dramatically reduced.
The variation ($\delta A_h$)
of $A_h$ in the range $1<\bar{g}_A<10$ is of order of $\delta A_h \sim 12$\%, 
corresponding to $A_h=28$\% and $A_h=25$\%,
for $\bar{g}_A=1$ and $\bar{g}_A=10$ respectively. The range of 
this variation should be
interpreted as the theoretical uncertainty of our scenario
on the inclusive observables,  at fixed value of $\Lambda=1$ TeV, that
becomes smaller by taking larger values of $\Lambda$.
In the case of the total cross sections, 
this deviation is even
smaller, being of the order of 0.8\%, which is a very negligible 
effect in comparison to the other (QCD and PDF) uncertainties affecting the 
strong interactions induced cross sections at the LHC.

Moreover, the results in Fig.\ref{fig8} are 
not very sensitive to the choice of the parametrization function. 
For instance, for $\bar{g}_A=1$,
the difference for $A_h$ evaluated by using $G_F(x)$ or 
$G_{\theta}(x)$ is of order of 8\%.

In conclusion, we believe that this scenario and in particular the 
$\Lambda=1-1.3$ TeV region that is required to explain the Tevatron anomaly, 
is still consistent with the inclusive measurements of $A_h$ reported by
the CMS and ATLAS collaborations within 2 standard deviations. 
This suggests that a dedicated 
experimental analysis of the $m_{tt}$ distributions of $A_h$ by the CMS and 
ATLAS collaboration is needed in order to either confirm or ruled out this scenario.

On the right plot of Fig.\ref{fig8} we show the corresponding 
predictions for the $m_{tt}$ inclusive observable $A_{LR}$. 
We can see that the general trend 
is an increase of the $A_{LR}$ values by increasing $\bar{g}_A$.
We did not show the SM prediction in the plot since this is about
one order of magnitude smaller.
We can see that the 
$A_{LR}$ approaches to a constant value for $4<\bar{g}_A<10$, namely 
$A_{LR}=7.4$\%. The variation of $A_{LR}$ in the considered  range of 
$\bar{g}_A$,  is of the order of $28$\%, passing from 5.8\% to 7.4\%.
Therefore, measurements of $A_{LR}$ at the LHC, even if inclusive in $m_{tt}$,
could be crucial for testing this model,
although a dedicated analysis of the $m_{tt}$ spectrum would be more 
effective and less model-dependent in constraining this scenario.

\section{Conclusions}
We have analyzed the impact of the gluon effective axial-vector 
coupling on the spin correlations $A_h$ and LR spin asymmetry 
$A_{LR}$ in top- antitop-quark production at the LHC. We studied these
observables at different invariant masses of the $t\bar{t}$ system and
showed that it would be necessary to measure these quantities as function of
the $t\bar{t}$ invariant mass $m_{tt}$ at the LHC. In particular, we 
found that these 
observables are very sensitive to the NP scale $\Lambda$ associated with 
the effective axial-vector 
coupling of gluon, in the high $t\bar{t}$ invariant mass regions
close enough to the scale $\Lambda$. 
Moreover, we found that the $A_{LR}$ is the best probe
to test this scenario at the LHC since the SM background is negligible.

We estimated the potential effect of the gluon effective axial-vector 
coupling on the $m_{tt}$ inclusive 
spin correlations measurements obtained by ATLAS and CMS collaboration.
We show that this scenario, for a scale $\Lambda \ge 1$ TeV, is 
still consistent with present measurements within standard deviations.
Therefore, a more dedicated analysis of those quantities as a function 
of $m_{tt}$ is mandatory in 
order to test this scenario at the LHC. We stress that the
8 TeV LHC has enough sensitivity either to confirm the 
Tevatron top charge asymmetry anomaly or to rule it out in the context 
of the considered NP scenario.

%%%%%%%%%%%%%%%%%%%%%%%%%%%%%%%%%%%%%%%%%%

\vspace{1cm}
\vbox{
\noindent{{\bf Acknowledgements} } \\
\noindent
We acknowledge useful discussions with W. Bernreuther and Z.-G. Si.
E.G. would like to thank the PH-TH division of CERN for its kind 
hospitality during the preparation of this work.
This work was supported by the ESF grants  8090, 8499, 8943, MTT8, MTT59, MTT60, MJD140, JD164, MJD298, by the recurrent financing SF0690030s09 project
and by  the European Union through the European Regional Development Fund.
}

\appendix

\section{Matrix elements}

Here we give the matrix elements for all possible helicity configurations of the initial and final state particles in partonic processes
\bea
q(k)\, \bar{q}(k^{\prime})&\to& t(p_t)\,  \bar{t}(p_{\bar{t}})\, ,
\\ \nonumber
\\ 
g(k)\, g(k^{\prime})&\to& t(p_t)\,  \bar{t}(p_{{\bar{t}}})\, ,
\eea
where $k,(k^{\prime})$ and $p_t,(p_{\bar{t}})$ denote  4-momenta of the quark (antiquark) or gluon (gluon) initial  and top- (antitop)-quark final states, respectively. The calculations were performed in the zero momentum frame (ZMF), where the $z$-axis was chosen in the direction of the top and all other momenta are assumed to lie on the $xz$-plane. In this frame the momenta 4-vectors for the top and antitop are 
\bea
p_{t} =	\frac{\sqrt{s}}{2}(1,0,0,\beta), \qquad p_{\bar t}	=	\frac{\sqrt{s}}{2}(1,0,0,-\beta), 
\eea
where $\hat s = (p_{\bar t} + p_{t})^{2}$ and $\beta = \sqrt{1 - 4 m_t^{2}/s}$. 
We compute the matrix elements for all possible helicity configurations of the initial and final particles. The spinors of helicity eigenstates are constructed by the helicity prescription, where the spin is given in the rest frame of the particle. The state is then boosted in the positive direction of the $z$-axis and then rotated clockwise in the $xz$-plane to end up with the chosen 4-momentum of the particle in the ZMF frame. 

The cross-section is given by
\bea
\frac{\mathrm{d}\sigma^{i}}{\mathrm{d}\Omega} &=& 
\frac{\beta}{4 \hat s} \alpha_S^2 c^{i} |\tilde{\mathcal{M}}^{i}|^{2}\, ,
\label{eq:Msquared}
\eea
where $i \in \{q\bar q, gg\}$, $c^{i}$ is an overall group theoretic factor and $|\tilde{\mathcal{M}}^{i}|^{2}$ is a non-normalized color averaged squared amplitude for the process. It can be expressed as
\bea
|\tilde{\mathcal{M}}^{i}|^{2} &=& 
\rho_{hh'}\bar\rho_{\bar h \bar h'} R^{i}_{hh',\bar h \bar h'}\, .
\label{eq:tildeM=R}
\eea
Here $R$ describes the production of on-shell top quark pairs from a given initial state. The matrices $\rho$, $\bar\rho$ are the density matrices describing the measurement of polarized top and antitop quarks in specific final states. The subscripts $h$ and $\bar h$ in Eq. \eqref{eq:tildeM=R} denote the top and antitop helicities. In the chosen basis for spin states $\rho = (1 + n^{i}_{t}\sigma_i)/2$ and $\bar\rho =(1 + n^{i}_{t}\sigma_3\sigma_i\sigma_3)/2$, where $\sigma_i$ are Pauli matrices. The corresponding covariant spin vectors are
\begin{align}
s_{t} = (\gamma\beta\, n_{t3},n_{t1},n_{t2},\gamma\,n_{t3}), 
\qquad
s_{\bar t} = (\gamma\beta\, n_{\bar t3},-n_{\bar t1},n_{\bar t2},-\gamma\,n_{\bar t3}),
\end{align} 
with $\gamma = \sqrt{s}/2m_t$. Helicity eigenstates correspond to $\vec{n} = (0,0,h)$ for both top and antitop, where $h$ is the sign of helicity. It takes values $+1$ and $-1$ denoting right-handed and left-handed fermions, respectively.

\subsection{Polarized $q\bar{q} \to t\bar t$ process}

The group theoretic factor for this process is
\bea
c^{q\bar{q}} &=& \frac{1}{4d(F)^{2}} d(A) \, =\,  \frac{N^{2}-1}{4 N^{2}},
\eea
where $d(F) = N$ and $d(A) = N^{2}-1$ are the dimensions of the fundamental (F) 
and adjoint (A) representation, respectively. 
The momenta of the initial quark and antiquark are	
\bea
k &=&\frac{\sqrt{s}}{2}(1,-\sin(\theta)\beta_q,0,\cos(\theta)\beta_q),
\qquad
k^{\prime} \,=\,  \frac{\sqrt{s}}{2}(1,\sin(\theta)\beta_q,0,-\cos(\theta)\beta_q), 
\eea
where $\theta$ the is the angle between the momenta of the initial quark and top in the ZMF and $\beta_{q} = \sqrt{1 - 4 m_q^{2}/s}$. 

The squared matrix element is given by \eqref{eq:Msquared}. For the initial $q\bar q$ the production matrix for a given initial state is
\bea
	R^{q\bar q}_{hh',\bar h \bar h';h_{q}h_{\bar q}}
	&	= \tilde{\mathcal{M}}^{q\bar q*}_{h'\bar h';h_{q}h_{\bar q}}\tilde{\mathcal{M}}^{q\bar q}_{h\bar h;h_{q}h_{\bar q}},
\eea
where
\bea
\tilde{\mathcal{M}}^{q\bar q}_{h\bar h;h_{q}h_{\bar q}}
&=& \delta_{h,\bar h}\delta_{h_{q},h_{\bar q}}	
\gamma_{q}^{-1}\gamma^{-1} \cos(\theta)	
\nonumber\\	
&+& \delta_{h,\bar h}\delta_{h_{q},-h_{\bar q}}
\gamma^{-1} \left(1 + h_{q} g_A \beta_q\right)  (-h_{q})\sin(\theta)
\nonumber\\
&+& \delta_{h,-\bar h}\delta_{h_{q},h_{\bar q}}
\gamma_{q}^{-1}\left(1 + h g_A \beta\right) (+h)\sin(\theta)
\nonumber\\
&+& \delta_{h,-\bar h}\delta_{h_{q},-h_{\bar q}}
\left(1 + h_{q} g_A \beta_q\right)\left(1 + h g_A \beta\right) (1 + h_{q}h\cos(\theta)),
\eea
where $\beta = \sqrt{1 - 4 m_t^{2}/s}$, 
$\gamma = \sqrt{s}/2m_t$, and $\gamma_{q} = \sqrt{s}/2m_q$.

After taking the spin sum over initial polarizations, the squared matrix element can be given by:
\bea
\frac{1}{4}\sum_{h_q,\bar{h}_q}|\tilde{\mathcal{M}}^{q\bar q}|^{2}  
&=& \mathcal{C}_{0}^{q\bar q}
+ n_{t}^{1}n_{\bar t}^{1}\mathcal{C}_{1}^{q\bar q}
+ n_{t}^{2}n_{\bar t}^{2}\mathcal{C}_{2}^{q\bar q}
+ n_{t}^{3}n_{\bar t}^{3}\mathcal{C}_{3} ^{q\bar q}
\nonumber\\
&+& (n_{t}^{1}n_{\bar t}^{3} + n_{\bar t}^{1}n_{t}^{3}) \mathcal{C}_{13}^{q\bar q}
+ (-n_{t}^{1} + n_{\bar t}^{1}) \mathcal{C}_{01}^{q\bar q}
+ (n_{t}^{3} - n_{\bar t}^{3}) \mathcal{C}_{03}^{q\bar q}\, ,
\eea
where
%
%\begin{subequations}
\bea
\mathcal{C}_0^{q\bar q}
&=& \frac{1}{8}
\Big(
2(1 + g_A^{2} \beta_q^{2})(1 + g_A^{2} \beta^{2}) 
+ (1 + g_A^{2} \beta_{q}^{2} + \gamma_{q}^{-2})(1 + g_A^{2} \beta^{2} + \gamma^{-2})
\nonumber\\
&+& 
\beta_{q}^{2}\beta^{2}(1 + g_A^{2})^{2} \cos(2\theta)
\Big)
+2 \beta_{q}\beta g_A^{2} \cos(\theta)\, ,
\\ 
\mathcal{C}_1^{q\bar q} 
&=& -\frac{1}{4}
\Big(\gamma_{q}^{-2}\gamma^{-2} 
+ \beta_{q}^{2}(1 + g_A^{2})(1 - g_A^{2} \beta^{2} + \gamma^{-2})\sin^2(\theta)
\Big),
\\
\mathcal{C}_2^{q\bar q}
&=&  -\frac{1}{4}
\Big(
-\gamma_{q}^{-2}\gamma^{-2}
+ \beta_{q}^{2}(1 + g_A^{2})(1 - g_A^{2} \beta^{2} - \gamma^{-2})\sin^2(\theta)
\Big)\, ,
\\
\mathcal{C}_3^{q\bar q}
&=&  -\frac{1}{8}
\Big(2(1 + g_A^{2} \beta_q^{2})(1 + g_A^{2} \beta^{2}) 
+ (1 + g_A^{2} \beta_{q}^{2} + \gamma_{q}^{-2})\beta^{2}(1 + g_A^{2})
\nonumber\\
&+&
\beta_{q}^{2}(1 + g_A^{2})(1 + g_A^{2} \beta^{2} + \gamma^{-2})\cos(2\theta)
\Big)
-2 \beta_{q}\beta g_A^{2} \cos(\theta)\, ,
\\
\mathcal{C}_{13}^{q\bar q}
&=& \gamma^{-1}\beta_{q}
\Big(\beta g_A^{2}
+\frac{1}{2}\beta_{q} (1 + g_A^{2})\cos(\theta)
\Big) \sin(\theta)\, ,
\\
\label{eq:Cqq01}
\mathcal{C}_{01}^{q\bar q}
&=& \gamma^{-1}\beta_{q} g_A 
\Big(1+	\frac{1}{2}\beta_{q} \beta(1 + g_A^{2})\cos(\theta) \Big) \sin(\theta)\, ,
	\\
\label{eq:Cqq03}
\mathcal{C}_{03}^{q\bar q}
&=&  g_A 
\left( \beta + \beta_{q} (1 + g_A^{2}\beta^{2})\cos(\theta)
+ \beta_{q}^{2}\beta \left(g_A^{2} - 
\frac{1}{2}(1+ g_A^{2})\sin^{2}(\theta)\right)
\right)\, .
\eea
%\end{subequations}
The coefficient $\mathcal{C}_0$ is proportional to the (final) spin summed result. The quotients $\mathcal{C}_i/\mathcal{C}_0$, $i \in \{1,2,3\}$ give spin correlations and the quotients $\mathcal{C}_i/\mathcal{C}_0$, $i \in \{01,03\}$ give the spin asymmetry for the corresponding quantization axis. Direction "3" corresponds to helicity. The term $\mathcal{C}_{03}^{q\bar q}$ is the only source of the spin asymmetry \eqref{ALR} at tree level.

The phase space integration is performed over the solid angle. The spin parameters $n^{1}_{t}$, $n^{2}_{t}$, $n^{1}_{t}$, $n^{2}_{t}$ are implicitly dependent on the azimuthal angle, so terms linear in these parameters vanish. Therefore the coefficients $\mathcal{C}_{01}$  and $\mathcal{C}_{13}$ do not contribute to the total cross-section. Only the sum $\mathcal{C}_{1} + \mathcal{C}_{2}$ is relevant after the phase space integration. In conclusion
\bea
\frac{1}{4\pi}\int  |\tilde{\mathcal{M}}^{q\bar q}|^{2} \,\mathrm{d} \Omega
&=& \mathcal{I}_{0}^{q\bar q} 
+ (n_{t}^{1}n_{\bar t}^{1}+n_{t}^{2}n_{\bar t}^{2})\, \mathcal{I}_{1+2}^{q\bar q}
+ n_{t}^{3}n_{\bar t}^{3}\, \mathcal{I}_{3} ^{q\bar q}
+ (n_{t}^{3} - n_{\bar t}^{3})\,  \mathcal{I}_{03}^{q\bar q}\, ,
\eea
where
%\begin{subequations}
\bea
\mathcal{I}_{0}^{q\bar q}
&=& \frac{1}{3} (1 + g_A^{2}\beta_q^{2} + \gamma_{q}^{-2}/2) (1 + g_A^{2}\beta^{2} + \gamma^{-2}/2)\, ,
\\
\mathcal{I}_{1+2}^{q\bar q}
&=& -\frac{1}{3} (1 + g_A^{2}\beta_q^{2} + \gamma_{q}^{-2}/2) (1 - g_A^{2}\beta^{2})\, ,
	\\
\mathcal{I}_{3}^{q\bar q}
&=& -\frac{1}{3} (1 + g_A^{2}\beta_q^{2} + \gamma_{q}^{-2}/2) (1 + g_A^{2}\beta^{2} - \gamma^{-2}/2)\, ,
\\
\mathcal{I}_{03}^{q\bar q}
&=& \frac{2}{3} (1 + g_A^{2}\beta_q^{2} + \gamma_{q}^{-2}/2) g_A\beta\, .
\eea
%\end{subequations}

\subsection{Polarized $gg \to t\bar t$ process}

The group theoretic overall factor for this process is
\bea
c^{gg} &=& \frac{d(F)C_F^{2}}{d(A)^{2}} \,=\, \frac{1}{4 N}\, ,
\eea
where $C_F = \frac{N^{2}-1}{2 N}$ is the quadratic Casimir invariant of the fundamental representation. The gluon momenta are $k=\frac{\sqrt{s}}{2}(1,-\sin(\theta),0,\cos(\theta))$ and $k^{\prime}= \frac{\sqrt{s}}{2}(1,\sin(\theta),0,-\cos(\theta))$, where $\theta$ the is the angle between gluon and top momenta in the ZMF.  The corresponding spin polarization vectors are $\epsilon_{\pm} = \frac{1}{\sqrt{2}}(1,\mp\sin(\theta),i,\mp \cos(\theta))$.

The production matrix takes a form
\bea
\label{eq:Rgg}
R^{gg}_{hh',\bar h \bar h';\lambda_g \lambda'_g}
& = 4
\begin{pmatrix}
\tilde{\mathcal{M}}^{tu}_{h'\bar h';\lambda_g \lambda'_g} \\
\tilde{\mathcal{M}}^{g}_{h'\bar h';\lambda_g \lambda'_g}
\end{pmatrix}^{\dagger}
\begin{pmatrix}
\mathcal{A} (\mathcal{A} - C_r)	& \mathcal{A} \beta \cos(\theta) C_r \\
\mathcal{A} \beta \cos(\theta) C_r & C_r \\
\end{pmatrix}
\begin{pmatrix}
\tilde{\mathcal{M}}^{tu}_{h\bar h;\lambda_g \lambda'_g} \\
\tilde{\mathcal{M}}^{g}_{h\bar h;\lambda_g \lambda'_g}\, ,
\end{pmatrix},
\eea
where $\mathcal{A} = (1 - \beta^{2}\cos^{2}(\theta))^{-1}$ and
\bea
C_r &=& \frac{C_{2}(G)}{4C_F} \,=\, \frac{N^{2}}{2(N^{2}-1)}\, ,
\eea
is a group theoretic constant, $0 \leq C_r \leq 1$, with $C_{2}(G) = N$ being the quadratic Casimir invariant in the adjoint representation. $C_r$ is independent of the normalization of the group generators and for Abelian groups $C_r = 0$. For abelian gauge theories $R$ is determined entirely by $\tilde{\mathcal{M}}^{tu}$, as one would expect. 

For gluon spins $\lambda_g$ and $\lambda'_g$ the amplitudes $\tilde{\mathcal{M}}^{tu}$ and $\tilde{\mathcal{M}}^{g}$ are
%
%\begin{subequations}
\bea
\tilde{\mathcal{M}}^{tu}_{h\bar h;\lambda_g \lambda'_g}	
&=& \delta_{h,\bar h} \delta_{\lambda_g,\lambda'_g}  \ \  
\gamma^{-1} \beta \sin^{2}(\theta)
\nonumber\\
&-& \delta_{h,\bar h} \delta_{\lambda_g,-\lambda'_g}  \
\gamma^{-1} (h\lambda_g + \beta)
\nonumber\\
&-& \delta_{h,-\bar h} \delta_{\lambda_g,\lambda'_g}  \
\beta(\lambda_g+h\cos(\theta))\sin(\theta),
\\
\tilde{\mathcal{M}}^{g}_{h\bar h;\lambda_g \lambda'_g}	
&=&\delta_{h,-\bar h} \delta_{\lambda_g,-\lambda'_g}\ 
g_A \beta \sin(\theta)\, .
\eea
%\end{subequations}
%
The axial coupling appears only in the non-abelian part $\tilde{\mathcal{M}}^{g}$ when top-quarks with opposite helicity are produced from gluons with opposite spin. The effect disappears for low energies and collinear momenta.

After taking the spin average over initial polarizations, the squared matrix element can be given in a relatively compact form:
\bea
\frac{1}{4}\sum_{\lambda_g,\lambda'_g}|\tilde{\mathcal{M}}^{gg}|^{2} 
&=& \mathcal{C}_{0}^{gg}  
+ n_{t}^{1}n_{\bar t}^{1}\mathcal{C}_{1}^{gg} 
+ n_{t}^{2}n_{\bar t}^{2}\mathcal{C}_{2}^{gg} 
+ n_{t}^{3}n_{\bar t}^{3}\mathcal{C}_{3}^{gg}  
\nonumber\\
&+& 
(n_{t}^{1}n_{\bar t}^{3} + n_{\bar t}^{1}n_{t}^{3}) \mathcal{C}_{13}^{gg} 
+ (-n_{t}^{1} + n_{\bar t}^{1}) \mathcal{C}_{01}^{gg}\, ,
\eea
where
%
%\begin{subequations}
\bea
\mathcal{C}_0^{gg}
&=& \mathcal{A} (\mathcal{A} - C_r) [1-\beta^{4}(1 + \sin^{4}(\theta))+ 2\beta^{2}\sin^{2}(\theta)] 
+ C_r g_A^{2} \beta^{2} \sin^{2}(\theta)\, ,
\\
\mathcal{C}_1^{gg}
&=& -\mathcal{A} (\mathcal{A} - C_r) [-\gamma^{-4} + (1 - \gamma^{-4})\sin^{4}(\theta)]
- C_r g_A^{2} \beta^{2} \sin^{2}(\theta)\, ,
\\
\mathcal{C}_2^{gg} 
&=& -\mathcal{A} (\mathcal{A} - C_r) [\gamma^{-4}+\beta^{4}\sin^{4}(\theta)] 
- C_r g_A^{2} \beta^{2} \sin^{2}(\theta)\, ,
\\
\mathcal{C}_3^{gg}
&=&  \mathcal{A} (\mathcal{A} - C_r) [1-\beta^{4}(1 + \sin^{4}(\theta))- 2\beta^{2}\sin^{2}(\theta)\cos^{2}(\theta)] 
- C_r g_A^{2} \beta^{2} \sin^{2}(\theta)\, ,
\\
\mathcal{C}_{13}^{gg}
&=&  \mathcal{A} (\mathcal{A} - C_r) \gamma^{-1}\beta^{2} \sin(2\theta)\sin^{2}(\theta)\, ,
\\
\mathcal{C}_{01}^{gg} 
&=&  C_r \mathcal{A} g_A \gamma^{-1}\beta^{3} \sin(2\theta)\, .
\eea
%\end{subequations}
The coefficient $\mathcal{C}_0$ is proportional to the (final) spin summed result and the rest are associated with different spin observables. Note that there is no LR-asymmetry for the $gg$-initial state, because there is no term similar to  Eq.\eqref{eq:Cqq03}. Instead, in this process the axial coupling introduces another strong spin asymmetry that is not present in the standard model. It is induced by the coefficient $\mathcal{C}_{01}^{gg}$ (and similarly by $\mathcal{C}_{01}^{q\bar q}$ \eqref{eq:Cqq01} for the $q\bar q$ initial state). This term is caused by the interference between the axial-vector and the vector couplings, and it is  the only term of this kind for the $gg$ initial  state. This term could induce azimuthal asymmetries. However, for the symmetric initial state, this effect averages out. In order to observe a physical azimuthal asymmetry induced by $\mathcal{C}_{01}^{gg}$, initial state polarization is needed.

The phase space averaged squared matrix element is given by
\bea
\frac{1}{4\pi}\int  |\tilde{\mathcal{M}}^{gg}|^{2} \,\mathrm{d} \Omega
&=& \mathcal{I}_{0}^{q\bar q} 
+ (n_{t}^{1}n_{\bar t}^{1}+n_{t}^{2}n_{\bar t}^{2})\, \mathcal{I}_{1+2}^{gg}
+ n_{t}^{3}n_{\bar t}^{3}\, \mathcal{I}_{3} ^{gg}\, ,
\eea
where
%\begin{subequations}
\bea
\mathcal{I}_{0}^{gg}
&=& -\frac{1}{16} \left(28 + 31 \gamma^{-2} - (32 + 32 \gamma^{-2} + 2 \gamma^{-4}) \frac{\alpha}{\beta}\right) 
+ \frac{3}{8}g_A^{2} \beta^{2}\, , 
\\
\mathcal{I}_{1+2}^{gg}
&=& -\frac{1}{16\beta^{2}} \left(10 + 23 \gamma^{-2} - \gamma^{-2} (32 + \gamma^{-2}) \frac{\alpha}{\beta}\right) 
- \frac{3}{8}g_A^{2} \beta^{2}\, ,
\\
\mathcal{I}_{3}^{gg}
&=& \frac{1}{16\beta^{2}} \left(60 - 25 \gamma^{-2} +  31 \gamma^{-4} - (32 + 4 \gamma^{-2} + 28 \gamma^{-4} + 2 \gamma^{-6}) \frac{\alpha}{\beta}\right) 
- \frac{3}{8}g_A^{2} \beta^{2}\, ,
\eea
%\end{subequations}
where $\alpha = \mathrm{atanh}(\beta)$ is the rapidity and the substitution $C_r = 9/16$ corresponding to SU(3) was made.

\bibliography{spinasym_final}

\end{document}